\documentclass[12pt]{article}
\usepackage{graphicx}
\advance\oddsidemargin by -\textwidth
\advance\oddsidemargin by -1in
\evensidemargin=\oddsidemargin
\def\fsl#1{\setbox0=\hbox{$#1$}           
   \dimen0=\wd0                                 
   \setbox1=\hbox{/} \dimen1=\wd1               
   \ifdim\dimen0>\dimen1                        
      \rlap{\hbox to \dimen0{\hfil/\hfil}}      
      #1                                        
   \else                                        
      \rlap{\hbox to \dimen1{\hfil$#1$\hfil}}   
      /                                         
   \fi}                                         %
\newcommand{\dfrac}[2]{\frac{\strut \displaystyle{#1}}
                      {\strut \displaystyle{#2}}}
\newcommand{\tr}{\mbox{tr}}

\newcommand{\hyper}[4]{F(#1,#2,#3; #4)}
\newcommand{\gtrsim}{\mathop{>}\limits_{\displaystyle{\sim}}}

\makeatletter
\def\@maketitle{\newpage
 \null
 {\normalsize \tt \begin{flushright} 
  \begin{tabular}[t]{l} \@date 
  \end{tabular}
 \end{flushright}}
 \begin{center}
 \vskip 2em
 {\LARGE \@title \par} \vskip 2.5em {\large \lineskip .5em
 \begin{tabular}[t]{c}\@author 
 \end{tabular}\par} 
 \end{center}
 \par
 \vskip 1.5em} 
\makeatother
\makeatletter
\@addtoreset{equation}{section}
\makeatother

\title{{\Large\bf
   Top Mode Standard Model with Extra Dimensions}}
\author{
  {\large
    Michio {\sc Hashimoto}\thanks{
      {\tt michioh@tuhep.phys.tohoku.ac.jp}}, 
    \quad
    Masaharu {\sc Tanabashi}\thanks{
      {\tt tanabash@tuhep.phys.tohoku.ac.jp}}
  }\\[5mm]
  {\it Department of Physics, Tohoku University}\\
  {\it  Sendai 980-8578, Japan}\\[1cm]
  {\large
    Koichi {\sc Yamawaki}\thanks{
      {\tt yamawaki@eken.phys.nagoya-u.ac.jp}}
  }\\[5mm]
  {\it Department of Physics, Nagoya University} \\ 
  {\it Nagoya 464-8602, Japan}
}
\date{
  TU-604 \\
  DPNU-00-34 \\
  hep-ph/0010260 \\
  October, 2000
}
\begin{document}
\maketitle
\begin{abstract}
We critically examine a version of the top mode standard model recently 
cast in extra dimensions by Arkani-Hamed, Cheng, Dobrescu, and Hall, 
based on  the (improved) ladder Schwinger-Dyson (SD) equation 
for the $D$-($=6,8$-)-dimensional gauge theories. 
We find that the bulk QCD cannot have larger coupling beyond the
{\em non-trivial ultraviolet (UV) fixed point}, the existence of which 
is supported by a recent lattice analysis.
The coupling strength at the fixed point is evaluated by using the
one-loop renormalization group equation.
It is then found that, in a version with only the third family 
(as well as the gauge bosons) living in the $D$-dimensional bulk, the 
critical (dimensionless) coupling 
for dynamical chiral symmetry breaking to occur 
is larger than the UV fixed point 
of the bulk QCD coupling 
for $D =6$, while smaller for $D=8$.  
We further find that the {\it improved} ladder SD equation in $D$ dimensions
has an {\it approximate 
scale-invariance} due to the {\it running} of the coupling and hence has
an {\it essential-singularity scaling} of the ``conformal phase transition,'' 
similar to Miransky scaling in 
the four-dimensional {\it ladder} SD equation with a {\it nonrunning} coupling.
This essential-singularity scaling can resolve the fine-tuning
even when the cutoff (``string scale'') is large.
Such a theory  has a 
{\it large anomalous dimension} $\gamma_m =D/2 - 1$
and is expected to be 
free from the flavor-changing-neutral-current problem
as in walking technicolor for $D=4$. Furthermore, the induced bulk 
Yukawa coupling
becomes finite even at infinite cutoff limit (in the formal sense), 
similarly to the renormalizability of the gauged Nambu-Jona-Lasinio model.
Comments are made on the use of the ``effective'' coupling,
which includes finite renormalization effects, instead of the 
${\overline {MS}}$ running coupling in the improved ladder SD equation.
\end{abstract}
\newpage

\section{Introduction}

The top quark condensate proposed by Miransky, Tanabashi and Yamawaki
(MTY)~\cite{MTYa,MTYb} and by Nambu~\cite{nambu} independently is a
natural idea to account
for the large mass of the top quark ($t$) on the weak-scale order in
contrast with other quarks and leptons. 
The Higgs boson in the standard model (SM) emerges as
a $\bar t t$ bound state and hence is closely connected with the top
quark itself. Thus the model may be called the ``top mode standard
 model (TMSM)''~\cite{MTYb}.

Actually, MTY introduced explicit four-fermion interactions~\cite{MTYa,MTYb}
\begin{equation}
  {\cal L}_{4f}
   = \frac{4 \pi^2}{N_c \Lambda^2} 
  \left[ 
   g_t (\bar\psi_L t_R)^2 + g_b (\bar\psi_L b_R)^2 
     + g^{(2)} \epsilon^{i,k} \epsilon^{j,l}
        (\bar\psi_L^i \psi_R^j)
         (\bar\psi_L^k \psi_R^l)
        + \mbox{h.c.}
         \right],
\label{4fermi}
\end{equation}
with  $i(j,k,l)=t, b$ for top and bottom quarks, 
 where $g_t, g_b, g^{(2)}$ are dimensionless four-fermion
couplings, $\Lambda$ is the cutoff, and $N_c$ is the number of colors,
and similarly for leptons as well as  the first and second generations of 
quarks and leptons.  While $g_t$ is responsible for the
top mass, the $g^{(2)}$ coupling is vital to the generation of the bottom mass 
without the problem of the axion.
MTY further  gave a concrete formulation based on the (improved) ladder 
Schwinger-Dyson (SD) equation for the QCD plus the
four-fermion interaction (\ref{4fermi}), the gauged Nambu--Jona-Lasinio
(NJL) model,
and found 
that when
 \begin{equation}
 g_t > g_{crit} >g_b
 \label{criticality}
 \end{equation}
  only the top quark can condense, giving
rise to the large top mass, while the bottom quark is kept massless, where 
$g_{crit}$ is
the critical coupling of the SD equation.
 As to the value of the top mass, MTY substituted the solution of the (improved) 
ladder SD equation into the Pagels-Stokar (PS) formula~\cite{PS} 
for $F_\pi = 250 \;{\rm GeV}$ and predicted $m_t \simeq 250 {\rm GeV}$ 
for the cutoff near the Planck scale~\cite{MTYa,MTYb}.

The model was further formulated in an elegant
fashion by Bardeen, Hill, and Lindner (BHL)~\cite{BHL} in the
SM language, 
based on the renormalization-group equation (RGE)
and the compositeness condition. 
This essentially incorporates $1/N_c$ subleading effects 
disregarded by the MTY paper.
The BHL model is in fact equivalent to the MTY model at $1/N_c$ leading 
order~\cite{Yama96}. 
Such $1/N_c$-subleading effects reduced the above MTY 
value 250 GeV to 220 GeV, a somewhat smaller value but still on the
order of the weak scale. 
Even this value, however, turned out to be a bit
larger than the mass of the top quark observed later .   

Quite recently, Arkani-Hamed, Cheng, Dobrescu, and Hall (ACDH)~\cite{ACDH} 
proposed a very 
interesting version of 
the TMSM in six and eight 
dimensions, in which the third family fermion and the gauge bosons are 
put in the $D$-($=6,8$-)-dimensional bulk, while the first and 
second families are in the 
four-dimensional brane (3-brane).
The model is largely based on the earlier papers~\cite{Dob,CDH},
which,
motivated by the topcolor~\cite{Hill91} and the top-seesaw
model~\cite{DH98}, 
proposed formulating the top quark condensate in the extra dimensions
in the spirit of large scale compactification scenarios~\cite{ADD,antoni}.
ACDH argued that the $D$-dimensional SM gauge couplings become strong 
due to Kaluza-Klein (KK) modes of the
standard model gauge bosons and hence may 
naturally give rise to the 
effective  four-fermion
interactions {\it in $D$-dimensional bulk} which have the same structure as  
Eq.(\ref{4fermi}), with $g_t > g_{crit} >g_b$, the situation 
similar to the original TMSM, Eq.(\ref{criticality}).~\footnote{
The previous studies in extra dimensions\cite{Dob,CDH,AMM} were
focused on the four-fermion interactions {\it in the 3-brane} in contrast to
those in the  ACDH model which are 
{\it in the $D$-dimensional bulk}~\cite{ACDH,kobak}.}
Moreover they argued that the top mass can be arranged to be a realistic 
value due to the effects of many KK modes of top quark even for the TeV
scale cutoff, thus the model may be free from serious  fine tuning 
as compared with the original TMSM having the cutoff near the Planck scale.

However, ACDH gave no dynamical 
arguments on whether dynamical symmetry breaking
really takes place or not in their model. They made an ansatz that
bulk strong gauge dynamics  in the ultraviolet
region near the cutoff (``string scale'') can well be 
simulated by the $D$-dimensional {\it bulk four-fermion
couplings} characterized by the cutoff scale.
They then calculated the {\it relative strength} of the bulk attractive forces 
among various channels based on the
most attractive channel (MAC) hypothesis~\cite{kn:RDS80}
and argued that only the top coupling
can be arranged to be above the critical coupling
like Eq.(\ref{criticality}) in the original mechanism of MTY. 
However, this would make sense only when these four-fermion
couplings were near the critical coupling, the  situation being what    
they simply assumed.
In fact, there is no information on the strength of
the bulk effective four-fermion couplings, which cannot be related in any
definite manner to the bulk gauge coupling, while the latter is  
calculable through matching with the low-energy SM coupling
in four dimensions (3-brane) at the compactification scale~\cite{DDG}.

In this paper, we shall study the dynamical issues of 
the ACDH version of the TMSM, 
based on the (improved) ladder SD equation
for the gauge theories 
in the bulk $D\;(=6,8)$ dimensions. 
As in ACDH~\cite{ACDH}, we here assume that the bulk anomaly 
may be cancelled by some stringy arguments like the Green-Schwarz mechanism.   
Then we present explicit solutions for the dynamical chiral symmetry
breaking (D$\chi$SB)
for $D$ dimensions with their implications on the ACDH scenario 
and reveal some salient features
of this dynamics for $D$ dimensions. 

We first discuss a {\em nontrivial ultraviolet (UV) fixed point} in
the one-loop renormalization-group equation of the ``truncated KK''
effective theory~\cite{DDG} of $D$-dimensional non-Abelian gauge
theories with compactified extra dimensions, in a manner similar 
to the analysis of $D=4+\epsilon$ ($0<\epsilon\ll 1$) gauge theories.
Although such a fixed point cannot be justified for $\epsilon \sim
{\cal O}(1)$ within the perturbative analysis, its existence is
supported by a recent lattice calculation\cite{EKM}.
Assuming the nonperturbative existence of such a fixed point, we then 
evaluate the gauge coupling strength at the fixed point by using
the one-loop RGE which was actually adopted by ACDH for their prediction
of the top quark mass. 
In the bulk SM, QCD is the only non-Abelian gauge
theory relevant to 
the D$\chi$SB. 
We then observe that {\it the $D$-dimensional bulk QCD coupling cannot
grow over the
fixed point value}, since  at a certain compactification scale
we match the bulk QCD coupling with 
 the 3-brane QCD coupling,  
which is obviously small, and hence the phase must 
be in the weak coupling regime below the fixed point.
The QCD coupling  for the ACDH version of the
TMSM  for $D=6,8$ is actually evaluated by the 
 truncated KK effective theory.

We next study the dynamical symmetry breaking in $D$-dimensional gauge
theories, based on the improved ladder SD equation~\cite{imp_ladder}, 
with
the $D$-dimensional bulk gauge coupling in the ladder SD equation being simply 
replaced by the [modified minimal subtraction scheme ${\overline {MS}}$] 
one-loop running coupling.
Actually,  in $D$-dimensional gauge theories
for both the fermion and the gauge bosons living in the $D$-dimensional bulk, 
with the extra dimensions being compactified, {\it 
dynamical symmetry breaking can be  triggered only 
by the dynamics in the ultraviolet region} where the 
gauge coupling becomes strong, and hence can be well described by the 
{\it $D$-dimensional} improved ladder SD equation, 
with {\it massless} gauge bosons {\it in the $D$-dimensional bulk},
{\it irrespectively of details of the infrared dynamics of the compactification
scale}. 

It is then found that, 
for the simplest version of the ACDH scenario with only the third family 
(as well as the gauge bosons) living in the bulk,
the UV fixed point 
of the bulk QCD coupling 
is smaller than the critical coupling 
for dynamical 
chiral symmetry breaking to occur for $D =6$,
while the situation is reversed
for $D=8$. 
That is {\em the dynamical symmetry breaking due to the bulk QCD 
dynamics cannot take place in
six dimensions and can in eight dimensions for the simplest ACDH 
version of the TMSM.}

Remarkably enough, the improved ladder SD equation with {\it running} coupling 
has an approximately {\it scale-invariant form}
 in $D$ dimensions and thus {\it the scaling
law is the essential-singularity type of ``conformal phase 
transition''}~\cite{MY} similar to Miransky scaling in the
four-dimensional 
ladder SD equation with {\it nonrunning} coupling~\cite{Miransky}.
Moreover, it has a {\it large anomalous dimension} $\gamma_m =D/2 -1$ near
the fixed point 
and hence has a chance to solve the flavor-changing-neutral current
problem as in walking technicolor for $D=4$~\cite{Hol}. 

This corresponds to a slowly damping mass function that 
still yields finiteness of the bulk decay constant $F_\pi^{(D)}$ and 
hence of the induced bulk Yukawa coupling even in the limit of infinite cutoff
(in the formal sense). 
Such a situation 
is similar to the renormalizability of the 
gauged NJL model
in four dimensions~\cite{KTY}.

We also comment that, instead of the ${\overline {MS}}$ running
coupling in the improved ladder SD equation,
we may use the ``effective'' coupling including  finite renormalization 
effects.
Unlike $\overline{MS}$ coupling, the ``effective'' gauge coupling
includes the effects of KK modes heavier than the renormalization
scale.
It is shown that the decoupling theorem is violated in the
``effective'' gauge coupling due to the summation of the large
number of KK modes.
Nevertheless, we find an {\it upper bound} on the effective gauge coupling
strength, which is roughly proportional to the UV fixed point in the
$\overline{MS}$ scheme.
We also show that the upper bound of the ``effective'' gauge coupling 
can be regarded as a UV fixed point of ``bare gauge coupling.'' 
Our results are therefore unchanged 
qualitatively even if we
adopt the ``effective'' coupling instead of $\overline{MS}$.

It should be emphasized, however, that finite renormalization can
affect our quantitative results, such as the value of the critical
coupling. 
The effective coupling tends to be stronger than 
${\overline {MS}}$ coupling and hence there appears the possibility that 
bulk SM couplings 
could lead to the top
quark condensate in the manner of Eq.~(\ref{criticality}) 
under certain conditions even for $D=6$ 
in the simplest ACDH version of
the TMSM. 

The paper is organized as follows.
In Section 2  we discuss the existence of the nontrivial UV fixed point 
in $D$-[$=(4+\epsilon)$-]-dimensional non-Abelian gauge theories in 
the $\epsilon$ expansion. 
Then we show the nontrivial UV fixed point in the 
$D$-[$=(4+\delta)$-]-dimensional non-Abelian gauge theories with the extra
$\delta\;(=2,4)$ dimensions 
compactified. The value of the UV fixed point for the ACDH version of the
TMSM is evaluated for $D=6,8$ based on the truncated KK effective
theory. We then give a rough argument why the ``strong''
 bulk QCD coupling  may not 
necessarily give rise to the condensate, based 
on a naive dimensional analysis (NDA)~\cite{NDA,kn:CLP00}.
In Section 3  
we derive  the $D$-dimensional ladder SD equation, and also
the improved ladder SD equation. In Section 4
we find numerically the critical values for 
D$\chi$SB to occur for  $D=6,8$, which are compared with
those of the ACDH couplings estimated in Section 2.  
The analytical solution is also 
obtained in further approximation,
and shows essential-singular-type scaling. In Section 5 we analyze the
operator product expansion (OPE) for the fermion propagator and identify the
anomalous dimension, which is then calculated to be $\gamma_m =D/2-1$.
In Section 6 the chiral fermion through the orbifold projection into the
3-brane is studied in some detail.
In Section 7 we discuss use of the effective coupling instead of
the ${\overline {MS}}$ running coupling in the improved ladder SD equation.   
Section 8 is devoted to the summary and discussion.
Appendix A contains formulas for the angular integration of the
ladder SD equation, which is a generalization of the previous 
result~\cite{kondo} 
to arbitrary (noninteger) $D$ dimensions.
Appendix B shows a gap equation of the NJL-type four-fermion 
model in $D\;(>4)$ dimensions, 
in which the scaling law is $1/g_{crit} -1/g \sim (m/\Lambda)^2$, 
essentially the
same (up to logarithms) as the NJL model for $D=4$, in sharp  
contrast to
the case of $D<4$ where the scaling law is given by 
$1/g_{crit} -1/g \sim (m/\Lambda)^{D-2}$~\cite{KY}.
Appendix C is for the approximation to the effective coupling discussed 
in Section 7.

\section{Existence of ultraviolet fixed point}
\label{sec:fixedpoint}

In order to illustrate the existence of the nontrivial ultraviolet fixed
point of the gauge theory in more than four dimensions, 
we start with a brief review of the gauge dynamics in
$D\;(=4+\epsilon$, $0<\epsilon\ll 1)$ dimensions~\cite{kn:Peskin80}
($\epsilon$ expansion).

The one-loop renormalization-group equation of the gauge
coupling is given by 
\begin{equation}
  \mu \dfrac{d}{d\mu} \hat g 
    = \frac{\epsilon}{2} \hat g + \dfrac{b_{(1)}}{(4\pi)^2} \hat g^3,
\label{eq:rge1}
\end{equation}
where $\hat g$ is the dimensionless gauge coupling scaled by a
renormalization scale $\mu$:
\begin{equation}
  g_D \equiv \dfrac{\hat g}{\mu^{\epsilon/2}},
\label{eq:dimless}
\end{equation}
with $g_D$ being the gauge coupling in $D\;(=4+\epsilon)$
dimensions and of 
mass dimension $-\epsilon/2$.
Hereafter we assume that the renormalization-group coefficient $b_{(1)}$ is
negative, $b_{(1)} < 0$.
{}From Eq.~(\ref{eq:rge1}), the renormalization-group flow of 
the dimensionless coupling $\hat g$ is given by
\begin{equation}
  \hat g^2(\mu) 
   = \dfrac{1}{
       \left(\frac{\mu'}{\mu}\right)^\epsilon \dfrac{1}{\hat g^2(\mu')}
      -\dfrac{2}{\epsilon}\dfrac{b_{(1)}}{(4\pi)^2} 
        \left[ 1 - \left(\frac{\mu'}{\mu}\right)^\epsilon \right]
     }.
\end{equation}
We then find an ultraviolet ($\mu\rightarrow\infty$) fixed point
\begin{equation}
  g_*^2 = \lim_{\mu\rightarrow\infty} \hat g^2(\mu)
        = \dfrac{\epsilon}{2} \dfrac{(4\pi)^2}{-b_{(1)}}.
  \label{eq:oneloop}
\end{equation}
It should be emphasized that $\epsilon$ is considered to be small here
and therefore the fixed point $g_*^2 \propto
\epsilon$ is still in its perturbative regime.
It is straightforward to extend the analysis to include higher-loop
effects.
The UV fixed point in the two-loop RGE, 
\begin{equation}
  \mu \frac{d}{d\mu} \hat g 
    = \frac{\epsilon}{2} \hat g 
     +\dfrac{b_{(1)}}{(4\pi)^2} \hat g^3
     +\dfrac{b_{(2)}}{(4\pi)^4} \hat g^5
\end{equation}
is given in terms of the $\epsilon$ expansion,
\begin{equation}
  g_*^2 = \dfrac{(4\pi)^2}{-b_{(1)}} \frac{\epsilon}{2} 
    \left(
       1 + \dfrac{ b_{(2)} }{b_{(1)}^2} \frac{\epsilon}{2} 
    \right) + {\cal O}(\epsilon^3).
  \label{eq:twoloop}
\end{equation}
The two-loop effect, $b_{(2)}$ term affects the coefficient of
$\epsilon^2$, keeping the coefficient of $\epsilon^1$ unchanged.
In fact, the $n$-loop effect can be regarded as an ${\cal
  O}(\epsilon^n)$  effect  in the $\epsilon$ expansion.
The perturbative stability of the fixed point $g_*$ is thus guaranteed
in the $D$-[$=(4+\epsilon)$-]-dimensional gauge theories.
We also note that the coefficients $b_{(1)}$ and $b_{(2)}$ are both
negative in QCD with $N_f \le 8$.
The two-loop UV fixed point Eq.~(\ref{eq:twoloop}) is thus smaller than 
the one-loop estimate Eq.~(\ref{eq:oneloop}).

Hence we expect that there exist (at least) two phases separated
by the fixed point $g_*$ in this theory. 
The weakly interacting phase $\hat g<g_*$ can be controlled
perturbatively.
It is therefore considered to be in the Coulomb phase
and the chiral symmetry is not broken in this phase.
On the other hand, the theory becomes strongly interacting 
in the low-energy region in the phase $\hat g > g_*$.
It is therefore expected to be in the confinement phase and the
chiral symmetry is expected to be broken dynamically.

Although the existence of such a fixed point for larger $\epsilon \sim 
{\cal O}(1)$ cannot be justified within perturbative analysis, 
recent analysis based on the lattice gauge theory~\cite{EKM} suggests that
the fixed point structure described above holds even at larger
(integer) values of $\epsilon$, if the extra dimensions are
compactified in a short distance.\footnote{
There still exists nontrivial phase structure even in the case of
noncompactified extra dimensions.
However, the phase transition is shown to be first order~\cite{kawai} and
we cannot obtain hierarchy between the cutoff scale and the low-energy scales.}
Actually, as we will see in the following, 
there exists a close correspondence between the RGEs of
$\epsilon\ll 1$  and of the compactified extra dimensions even within
the perturbative approach.

Now we evaluate the {\it nontrivial UV fixed point of the gauge theory 
in $D\;(=4+\delta)$ dimensions where the extra $\delta$ dimensions are 
compactified}. In this case
we need to deal with an 
infinite number of KK modes
above the compactification scale $R^{-1}$.
However, the KK modes heavier than the renormalization scale $\mu$ are
actually decoupled in the RGE.
We only need to sum up the loops of KK modes lighter than $\mu$.
This approach is called ``truncated KK'' effective theory~\cite{DDG}.
The theory can be fully controlled in this truncated KK effective 
theory. 

The RGE of the gauge coupling ($g$) on the 3-brane is given by
\begin{equation}
  (4\pi)^2 \mu\dfrac{d}{d\mu} g = N_{\rm KK} b' g^3
\label{eq:rge2}
\end{equation}
in the truncated KK effective theory.
Here $N_{\rm KK}$ stands for the number of KK modes below the
renormalization scale $\mu$.
The RGE factor $b'$ is given by
\begin{equation}
  b' = -\dfrac{26-D}{6} C_G + \dfrac{\eta}{3} T_R N_f,
  \label{RGcoef}
\end{equation}
where $\eta$ represents the dimension of the spinor representation of 
$SO(1,D-1)$,
\begin{equation}
  \eta \equiv \tr_\Gamma 1 = 2^{D/2} \qquad
  \mbox{for even $D$},
\end{equation}
and $N_f$ is the number of fermions in the bulk.
The group-theoretical factors $C_G$ and $T_R$ are given by
$C_G=N$ and $T_R=1/2$ for $SU(N)$ gauge theory.

For sufficiently large $\mu\gg R^{-1}$, $N_{\rm KK}$ is estimated as~\cite{DDG}
\begin{equation}
  N_{\rm KK} = \frac{1}{n} 
               \dfrac{\pi^{\delta/2}}{\Gamma(1+\delta/2)} (\mu R)^\delta,
\end{equation}
where we have assumed that the extra dimensions are compactified to
an orbifold $T^\delta/Z_n$ with $Z_n$ being a discrete group with order of 
$n$.

The gauge coupling on the 3-brane can be related to 
the gauge coupling in the $D$-dimensional bulk $g_D$ as
$g^2_D = (2\pi R)^\delta g^2/n$. 
Thus the {\it dimensionless} bulk gauge coupling $\hat g$ can be
defined following Eq.~(\ref{eq:dimless}):  
\begin{equation}
  \hat g^2 = \dfrac{(2\pi R\mu)^\delta}{n} g^2.
\label{eq:dimless2}
\end{equation}
Substituting Eq.~(\ref{eq:dimless2}) into Eq.~(\ref{eq:rge2}), we obtain
\begin{equation}
  \mu \dfrac{d}{d\mu} \hat g
  = \frac{\delta}{2} \hat g 
   +(1+\delta/2) \Omega_{\rm NDA} b' \hat g^3,
\label{eq:rge1a}
\end{equation}
with $\Omega_{\rm NDA}$ being the loop factor in the 
NDA~\cite{NDA,kn:CLP00}:
\begin{equation}
  \Omega_{\rm NDA} \equiv \dfrac{1}{(4\pi)^{D/2} \Gamma(D/2)}, \qquad
  D = 4 + \delta.
\end{equation}
It is interesting to note a similarity between Eq.~(\ref{eq:rge1})
and Eq.~(\ref{eq:rge1a}):
The factor $\epsilon$ in the $\epsilon$ expansion corresponds to 
$\delta$ in the truncated KK effective theory with the simple replacement
of $b/(4\pi)^2$ by
$(1+\delta/2)\Omega_{\rm NDA}b'$. 

The RGE Eq.~(\ref{eq:rge1a}) can easily be solved as
\begin{equation}
  \hat g^2(\mu) 
   = \dfrac{1}{
       \left(\frac{\mu'}{\mu}\right)^\delta \dfrac{1}{\hat g^2(\mu')}
      -\left(\dfrac{2}{\delta}+1\right) \Omega_{\rm NDA} b'
        \left[ 1 - \left(\frac{\mu'}{\mu}\right)^\delta \right]
     }.
     \label{ghat}
\end{equation}
Thus, we find a {\it nontrivial UV fixed point}
\begin{equation}
  g_*^2 \Omega_{\rm NDA} 
  = \dfrac{1}{-\left(\frac{2}{\delta} + 1\right) b'}.
\label{eq:fixed}
\end{equation}
It should be noted that the coupling $\hat g^2$ in Eq.~(\ref{ghat})
grows very quickly close to the value of the fixed point.
  
In the ACDH scenario~\cite{ACDH} of the TMSM, the top quark
interaction responsible 
for the dynamical electroweak symmetry breaking 
is assumed to come
(mainly) from the bulk QCD interaction.
On the other hand, the low-energy QCD coupling in the 3-brane 
is obviously well below its fixed point.
[$\hat g^2/g_*^2 \simeq 1.83\alpha/n$ ($0.72\alpha/n$) with
$\alpha\equiv g^2/(4\pi)$ for $D=6$ ($D=8$) at $\mu=R^{-1}$. ]
Thus, Eq.~(\ref{eq:fixed}) can be regarded as {\it the upper bound of the 
dimensionless coupling of the bulk QCD\@}.
In fact, in the ACDH scenario for $D=6$ 
the upper bound of the dimensionless QCD coupling
is given by
\begin{equation}
  C_F \hat g^2 \Omega_{\rm NDA}< C_F g_*^2 \Omega_{\rm NDA}
  \simeq 0.09, 
\label{eq:upper}
\end{equation}
where we have used $\delta=2$,  $\eta=8$,  $C_F=4/3$,  $C_G=3$, and
$N_f=2$ in Eqs.~(\ref{RGcoef}) and (\ref{eq:fixed}).
Even though the value of $g_*$ can be affected by the higher-loop
effects, it is proportional to $1/(-b')$ for sufficiently large
$-b'$.\footnote{
It should be emphasized, however, that the higher-loop effects cannot be
made arbitrarily small even in the large $-b'\;(>0)$ limit.
Equation (\ref{RGcoef}) shows that large $-b'$ corresponds to
large $C_G$ (and small $N_f$) with $C_G=N$ for $SU(N)$ gauge theory.
The typical size of the $n$-loop effect at the fixed point is thus of
order $(N g_*^2 \Omega_{\rm NDA})^n \sim [N/(-b')]^n \simeq
[6/(26-D)]^n$ even in the large $N$ limit.}
Equation (\ref{eq:upper}) implies that, 
although the bulk QCD coupling is generally
expected to become ``strong'' in the region beyond the compactification scale,
it is actually {\it not so strong} as to make the perturbative expansion totally 
useless.
Moreover, an analysis similar to Eq.~(\ref{eq:twoloop}) indicates that
the value of $g_*$ of the bulk QCD with $\delta=2$ and $N_f=2$ tends
to be decreased by taking into account the two-loop effects.
The estimate Eq.~(\ref{eq:upper}) can thus be regarded as a conservative
one, although we expect sizable higher-loop uncertainty in our
estimate.

It is therefore quite {\it  nontrivial whether bulk QCD can be strong
enough to trigger dynamical electroweak symmetry breaking}. 
Before starting a detailed analysis, it would be helpful to give a
simpler discussion from the viewpoint of the NDA~\cite{NDA,kn:CLP00} and
MAC~\cite{kn:RDS80}, which leads to the
condition for the dynamical symmetry breaking to take place 
\begin{equation}
  C_F \hat g^2 \Omega_{\rm NDA} \gtrsim 1,
\end{equation}
where $\Omega_{\rm NDA}$ comes from the loop suppression factor of NDA\@ and
$C_F$ is the quadratic Casimir of the fundamental representation, which 
is from the MAC assumption.
Hence Eq.~(\ref{eq:upper}) suggests that bulk QCD may not be enough to induce
dynamical electroweak symmetry breaking in the ACDH scenario in 
six dimensions.

However, the
present analysis might be too simple minded.
In the following sections we will investigate this issue using the 
SD gap equation within the improved ladder approximation.

\section{Improved ladder Schwinger-Dyson equation}
\subsection{Ladder SD equation}

We next investigate the condition for the chiral symmetry to break dynamically 
in gauge theories in dimensions $D>4$.
The dimensions of the space-time $D$ need to be even in order that
the chiral symmetry is defined in the bulk.
Since the electroweak symmetry is a chiral symmetry, the condition
studied in this section can be regarded as the condition for 
dynamical electroweak symmetry breaking in the bulk.

Bulk D$\chi$SB in gauge theories with extra dimensions is  
considered as a nonperturbative effect of the high-energy 
region~\cite{ACDH} where the SM gauge couplings in the $D$-dimensional bulk
become strong. 
We therefore neglect the infrared dynamics due to 
the finite size effects of extra dimensions in
the following.

\begin{figure}[t]
  \begin{center}
    \includegraphics{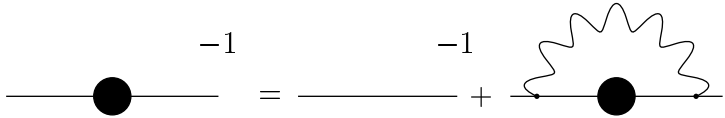}
    \caption{
      Feynman diagram of the SD equation in the ladder approximation.  
      Solid lines with and without a blob represent dressed and
      bare propagators of fermions [$S(p)$, $S_0(p)$], respectively. 
      The gauge boson-propagator ($D_{MN}$) is denoted by a wavy line.
      \label{fig:SD_graph}}
  \end{center}
\end{figure}

The $D$-dimensional ladder SD equation for the fermion propagator is given by
Fig.~\ref{fig:SD_graph}.
It then reads
\begin{equation}
  i S^{-1}(p) = i S_0^{-1}(p)
   +\int \!\! \dfrac{d^D q}{(2\pi)^D i} 
   \left[
     -ig_D T^a \Gamma^M \right] S(q) \left[ -ig_D T^a \Gamma^N
   \right] D_{MN}(p-q),
\label{SDeqD}
\end{equation}
where $S$ and $S_0$ denote dressed and bare propagators of the fermion,
respectively.
Within the ladder approximation, 
the gauge boson propagator $D_{MN}$ is approximated 
at the  tree level 
by the form 
\begin{equation}
   D_{MN}(p-q) =
   \dfrac{-i}{(p-q)^2} \left[
     g_{MN} - (1-\xi) \dfrac{(p-q)_M (p-q)_N}{(p-q)^2}
   \right],
\end{equation}
with $\xi$ being a gauge-fixing parameter.
We also indicate the the gamma matrix of $SO(1,D-1)$ by $\Gamma^M$ 
\begin{equation}
  \{ \Gamma^M, \Gamma^N \} = 2 g^{MN}, 
  \qquad
  M,N=0,1,2,3,5, \cdots, D.
\end{equation}

Since we are dealing with D$\chi$SB in the bulk, we take the bare propagator 
of the fermion in $D$ dimensions to be massless, $i S_0^{-1}(p) = \fsl{p}$.
The dressed propagator $S$ may be written as 
\begin{equation}
  iS^{-1}(p) = A(-p^2) \fsl{p} - B(-p^2).
\end{equation}
Then Eq.~(\ref{SDeqD}) leads to coupled SD equations after Wick rotation:
\begin{eqnarray}
  A(p_E^2)
    &=& 1 + \dfrac{C_F g_D^2}{p_E^2} \int \!\! \dfrac{d^D q_E}{(2\pi)^D}
            \dfrac{A(q_E^2)}{A^2 q_E^2 + B^2}
    \nonumber\\
    & & \hspace{-2em} \times
            \left[
              -(3-D-\xi) \dfrac{ p_E \cdot q_E}{(p_E-q_E)^2}
              +2 (1-\xi) \dfrac{p_E\cdot (p_E-q_E) q_E\cdot (p_E-q_E)}
                               {(p_E-q_E)^4}
            \right], 
\label{eq:sda}
            \\
  B(p_E^2)
    &=& (D-1+\xi) C_F g_D^2 \int \!\! \dfrac{d^D q_E}{(2\pi)^D}
            \dfrac{B(q_E^2)}{A^2 q_E^2 + B^2} 
            \dfrac{1}{(p_E-q_E)^2},
\label{eq:sdb}
\end{eqnarray}
where $p_E$ and $q_E$ denote the Euclidean momenta 
$p_E^2 \equiv -p^2$, $q_E^2 \equiv -q^2$, respectively.
Performing the angular integrals in Eq.~(\ref{eq:sda}) and
Eq.~(\ref{eq:sdb}), we find 
\begin{eqnarray}
  A(x) &=& 1 + 2\dfrac{D-2}{D} \xi
               \Omega_{\rm NDA}
               \dfrac{C_F }{x}
               \int_0^{\Lambda^2} dy y^{D/2-1}
               \dfrac{g_D^2 A(y)}{A^2y+B^2}K_A(x,y),
\label{eq:sda3}
       \\
  B(x) &=& (D-1+\xi) \Omega_{\rm NDA} C_F 
           \int_0^{\Lambda^2} dy y^{D/2-1} 
           \dfrac{g_D^2 B(y)}{A^2y+B^2}K_B(x,y),
\label{eq:sdb3}
\end{eqnarray}
with $x\equiv p_E^2$, $y\equiv q_E^2$,
where we have introduced the ultraviolet cutoff $\Lambda$, which is believed 
to have physical meaning such as the string scale in this class of
models with extra dimensions.

The integral kernels $K_A$ and $K_B$ are given in Ref.~\cite{kondo}
and are explicitly written as
\begin{eqnarray}
  K_A(x,y) &=& \dfrac{y}{x}\theta(x-y) + (x \leftrightarrow y), \\
  K_B(x,y) &=& \dfrac{1}{x}\theta(x-y) + (x \leftrightarrow y)
\end{eqnarray}
for $D=4$, 
\begin{eqnarray}
  K_A(x,y) &=& 
    \frac{y}{x}\left(1-\frac{y}{2x}\right)\theta(x-y) + (x \leftrightarrow y),
  \\
  K_B(x,y) &=& 
    \frac{1}{x}\left(1-\frac{y}{3x}\right)\theta(x-y) + (x \leftrightarrow y)
\end{eqnarray}
for $D=6$, and
\begin{eqnarray}
  K_A(x,y) &=& 
    \frac{y}{x}\left(1-\frac{4y}{5x}+\frac{y^2}{5x^2}\right)\theta(x-y)
    + (x \leftrightarrow y),
  \\
  K_B(x,y) &=& 
    \frac{1}{x}\left(1-\frac{y}{2x}+\frac{y^2}{10x^2}\right)\theta(x-y)
    + (x \leftrightarrow y)
\end{eqnarray}
for $D=8$. (See Appendix \ref{sec:app1} for details.)

Hereafter, we will use the Landau gauge $\xi=0$ in which the wave
function renormalization is absent [$A(x)\equiv 1$] within the ladder
approximation.

\subsection{Improved ladder SD equation}

It should be recalled here that we have so far neglected effects
of the running of the gauge coupling.
The powerlike behavior of the running coupling makes its
effects extremely important, however.
In the analysis of D$\chi$SB in four-dimensional gauge theories, a
widely used 
approximation is the so-called 
``improved'' ladder approximation~\cite{imp_ladder},
in which the renormalization point $\mu^2$ of the running coupling
constant in the SD equation is replaced by $\max(p_E^2, q_E^2)$.
This is a successful approximation for explaining properties of low energy
QCD phenomenology.
In the following analysis we adopt the improved ladder approximation
and replace the gauge coupling  $g_D^2$ in Eq.~(\ref{eq:sdb3})
by
\begin{equation}
 g_D^2 \rightarrow g_D^2(p_E,q_E) = 
   \dfrac{\hat g^2(|p_E|)}{(p_E^2)^{\delta/2}} \theta(|p_E|-|q_E|)
  +\dfrac{\hat g^2(|q_E|)}{(q_E^2)^{\delta/2}}\theta(|q_E|-|p_E|). \label{g_D}
\end{equation}

As we discussed in Section \ref{sec:fixedpoint}, the UV
fixed point $g_*$ plays the role of the upper bound of $\hat g$ in the 
ACDH scenario of the TMSM.
We also note that the {\it dimensionless} bulk gauge coupling  
$\hat g$ in Eq.~(\ref{ghat}) approaches
its fixed point very quickly for $\mu>1/R$ due to its power-law
running and hence is {\it near the fixed point value over a wide range of 
the momentum in the integral of the SD equation}. 
For determination of the condition of the bulk D$\chi$SB,
it is therefore sufficient to investigate the SD equation with the coupling 
just on the
UV fixed point:
\begin{equation}
  B(x) = (D-1) \kappa_D \int_{M_0^2}^{\Lambda^2} dy y^{D/2-1}
    \dfrac{B(y)}{y+B^2(y)} K_B^{\rm imp}(x,y) 
\label{eq:impsdeq}
\end{equation}
with
\begin{equation}
  \kappa_D \equiv C_F g_*^2 \Omega_{\rm NDA}
   = \dfrac{C_F}{
        \left(\frac{2}{D-4}+1\right) \left[
          \frac{26-D}{6} C_G - \frac{\eta}{3}T_R N_f
        \right]},
\label{eq:kappa_def}
\end{equation}
where we have used  Eqs.~(\ref{RGcoef}) and (\ref{eq:fixed}),
and
 $K_B^{\rm imp}$ is given by
\begin{equation}
  K_B^{\rm imp}(x,y) = 
    \frac{1}{x^2}\left(1-\frac{y}{3x}\right)\theta(x-y)
      + (x \leftrightarrow y)
  \qquad
  \mbox{for $D=6$}
\end{equation}
and
\begin{equation}
  K_B^{\rm imp}(x,y) = 
    \frac{1}{x^3}\left(1-\frac{y}{2x}+\frac{y^2}{10x^2}\right)\theta(x-y)
    + (x \leftrightarrow y)
  \qquad
  \mbox{for $D=8$}.
\end{equation}
Since the extra dimensions are compactified below the scale $1/R$, we
have introduced the infrared (IR) cutoff $M_0$ in Eq.~(\ref{eq:impsdeq}).
However, {\it the bulk D$\chi$SB becomes insensitive to $M_0$ for large
$\Lambda$} as we will show in the next section.
It is to be noted that the resulting {\it improved ladder} SD equation 
with {\it running} coupling in Eq.~(\ref{eq:impsdeq}) is 
a {\it scale-invariant} form, similar to the ladder SD equation with 
{\it constant} gauge coupling in four dimensions.

We note that, in the improved ladder
SD equation Eq.~(\ref{eq:impsdeq}), the $\kappa_D$ defined by
 Eq.~(\ref{eq:kappa_def})  plays the role
of a ``coupling.'' 
It can be shown that there exists a critical $\kappa_D$ above which
D$\chi$SB takes place for sufficiently large $\Lambda$ in the bulk.

\section{Analysis of the improved ladder SD equation}
\subsection{Numerical study}

\begin{figure}[t]
  \begin{center}
    \includegraphics{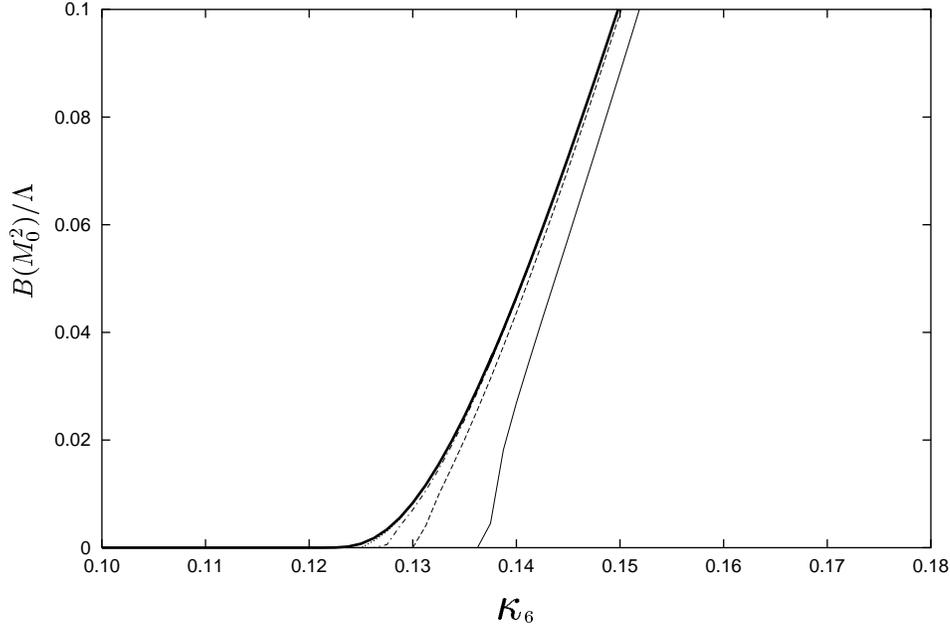}
    \caption{The scaling behavior in six dimensions.
             The lines from right to left are graphs for 
             $\Lambda^2/M_0^2=10^3,10^4,10^5,10^6,10^{10}$, respectively.
             \label{fig:cri_6dim}}
  \end{center}
\end{figure}

The aim of this section is to determine the critical $\kappa_D$ (and
the scaling behavior around $\kappa_D^{\rm crit}$) by solving the SD
equation Eq.~(\ref{eq:impsdeq}) in a numerical method.

Let us start with the case $D=6$ and consider a discretized version of 
Eq.~(\ref{eq:impsdeq}):
\begin{equation}
  B_i = 5\kappa_6 \left[
    \sum_{j=1}^i \dfrac{x_j B_j}{x_j + B_j^2}
                 \dfrac{x_j^2}{x_i^2}\left(1- \dfrac{x_j}{3x_i}\right)
   +\sum_{j=i+1}^{i_\Lambda} \dfrac{x_j B_j}{x_j + B_j^2}
                 \left(1- \dfrac{x_i}{3x_j}\right)
   \right]
\label{eq:discrete}
\end{equation}
with $i,j$ being integer indices and
\begin{equation}
  x_j \equiv M_0^2 \exp\left[
    \dfrac{j-1}{i_\Lambda -1 }\ln \dfrac{\Lambda^2}{M_0^2}
  \right], \qquad
B_j \equiv B(x_j).
\end{equation}
In order to solve the discretized SD equation Eq.(\ref{eq:discrete}),
a series $B_j^{(n)}$ is defined by a recursion relation,
\begin{equation}
  B^{(n+1)}_i \equiv 5\kappa_6 \left[
    \sum_{j=1}^i \dfrac{x_j B^{(n)}_j}{x_j + (B^{(n)}_j)^2}
                 \dfrac{x_j^2}{x_i^2}\left(1- \dfrac{x_j}{3x_i}\right)
   +\sum_{j=i+1}^{i_\Lambda} \dfrac{x_j B^{(n)}_j}{x_j + (B^{(n)}_j)^2}
                 \left(1- \dfrac{x_i}{3x_j}\right)
   \right]
\end{equation}
and the initial condition
\begin{equation}
  B_j^{(n=0)} = M_0 \qquad
  \mbox{for $j=1,2, ... \,, i_\Lambda$}.
\end{equation}
For sufficiently large $n$, the series $B_j^{(n)}$ is numerically
shown to converge to a certain $B_j$, which is nothing but the
solution of the SD equation Eq.(\ref{eq:discrete}).
It is also confirmed that the solution is insensitive to the value of
$i_\Lambda$, if $i_\Lambda$ is taken to be large enough.

Figure \ref{fig:cri_6dim} shows the scaling behavior of the order parameter 
of D$\chi$SB, $B(M_0)=B_{j=1}$, near the critical $\kappa_6$.
We find
\begin{equation}
  \kappa_6^{\rm crit} \simeq 0.122.
\label{eq:numcrit6}
\end{equation}
On the other hand, the $\kappa_6$ of the ACDH scenario with $D=6$ (QCD 
with two flavor fermions in the bulk) can be calculated from 
Eq.~(\ref{eq:kappa_def}).
We find
\begin{equation}
  \kappa_6^{\rm ACDH} = \frac{1}{11} \simeq 0.091,
\label{eq:ACDH6}
\end{equation}
where we have used $C_F=4/3$, $C_G=3$, $N_f=2$, and $\eta=8$.
Note that this is the {\it upper bound} of the bulk QCD coupling.
We therefore conclude that the simplest version of the ACDH scenario
does not work properly
in $D=6$ dimensions within the 
 improved ladder
approximation. 

\begin{figure}[t]
  \begin{center}
    \includegraphics{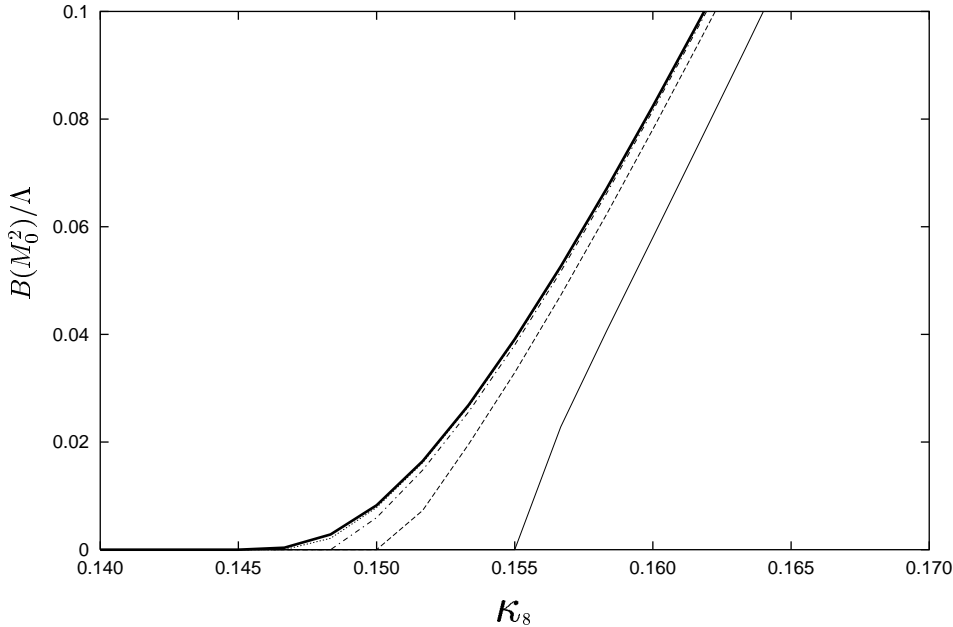}
    \caption{The scaling behavior in eight dimensions.
             The lines from right to left are graphs for 
             $\Lambda^2/M_0^2=10^3,10^4,10^5,10^6,10^{10}$, respectively.
             \label{fig:cri_8dim}}
  \end{center}
\end{figure}

A similar analysis is also performed for $D=8$ dimensions.
We obtain the scaling behavior of Fig.~\ref{fig:cri_8dim}, 
and the critical $\kappa_8$
\begin{equation}
  \kappa_8^{\rm crit} \simeq 0.146.
\label{eq:numcrit8}
\end{equation}
Since the ACDH scenario in $D=8$ dimensions predicts
\begin{equation}
 \kappa_8^{\rm ACDH} = \frac{8}{33} \simeq 0.242,
\label{eq:ACDH8}
\end{equation}
there is the possibility to construct viable models in $D=8$
within the 
improved ladder approximation.

One may doubt the validity of the ladder approximation in this model.
The size of nonladder corrections is estimated to be $1\%$--$20\%$ in 
the analysis of four-dimensional walking technicolor~\cite{kn:ALM88}.
We expect a similar size of nonladder corrections in the present model.
On the other hand, the fixed point Eq.~(\ref{eq:ACDH6}) is smaller 
than the critical value Eq.~(\ref{eq:numcrit6}) by more than $25\%$.
Although it is extremely difficult to draw a definite conclusion from
these numbers, it is likely that the ACDH scenario in $D=6$ dimensions 
is still in the chiral symmetric phase even in beyond-the-ladder approximations.
We also note that the ladder results are
qualitatively consistent with the naive dimensional analysis described 
in Section 2.
The bulk QCD coupling is not so strong as to destroy the
perturbative picture completely, anyway.
We thus expect that our results (Fig.~2 and Fig.~3) will be unchanged
qualitatively even beyond the ladder approximation.

\subsection{Analytical study}

The improved ladder SD equation can be investigated analytically by
applying further approximations.
The SD equation can be greatly simplified if the integral kernel
$K_B^{\rm imp}$ is approximated by
\begin{equation}
  \tilde K_B^{\rm imp}(x,y)
  = \dfrac{1}{x^{D/2-1}}\theta(x-y) + (x\leftrightarrow y).
\label{eq:appkern}
\end{equation}
The approximation Eq.~(\ref{eq:appkern}) can be justified in a wide
range of the integral region ($x \not{\!\!\simeq} \; y$) in
Eq.~(\ref{eq:impsdeq}).
We also note that the kernel Eq.~(\ref{eq:appkern}) has scale
invariance like the original kernel $K_B^{\rm imp}$.

Although the SD equation Eq.~(\ref{eq:impsdeq}) is still nonlinear even
under this approximation, we can overcome the difficulty by using the
bifurcation technique~\cite{kn:Atkinson}, in which the mass function
$B$ in the denominator in the SD equation is eliminated and an
infrared cutoff $M\equiv B(M^2)$ is introduced instead.
The bifurcation technique is justified when $\kappa_D$ is close to its 
critical point.

The SD equation Eq.~(\ref{eq:impsdeq}) then leads to a linear equation 
\begin{equation}
  B(x) = (D-1) \kappa_D \int_{M^2}^{\Lambda^2} dy y^{D/2-2}
               B(y) \tilde K_B^{\rm imp}(x,y)
\label{eq:appsdeq}
\end{equation}
and a subsidiary condition
\begin{equation}
  M = B(M^2).
\label{eq:subsidiary}
\end{equation}
The integral equation Eq.~(\ref{eq:appsdeq}) is equivalent to a set of
a differential equation,
\begin{equation}
  \left[
    x^2 \dfrac{d^2}{dx^2} + \frac{D}{2} x \dfrac{d}{dx}
      + (D-1)(D/2-1) \kappa_D \right] B(x) = 0,
  \label{eq:diffeq}
\end{equation}
and boundary conditions
\begin{eqnarray}
  & & 
  \hbox to 9cm{$\displaystyle
    \left. \dfrac{d}{dx} B(x) \right|_{x=M^2} = 0 
  $ \hfill (IR-BC) \hfil}
  \label{eq:irbc}
  \\
  & &
  \hbox to 9cm{$\displaystyle
    \left. \left[ x \dfrac{d}{dx}  - 2\omega \right] B(x) 
    \right|_{x=\Lambda^2} = 0 
  $ \hfill (UV-BC),\hfil }
  \label{eq:uvbc}
\end{eqnarray}
with $\omega$ being defined by
\begin{equation}
  \omega \equiv -\frac{1}{2} \left(\frac{D}{2} - 1 \right).
\end{equation}

It is easy to solve the differential equation Eq.~(\ref{eq:diffeq}).
Combined with the subsidiary condition Eq.~(\ref{eq:subsidiary}) and
the infrared boundary condition (IR-BC), we find
\begin{equation}
  \dfrac{B(x)}{M}
 = \frac{1}{2\tilde \nu} \left(\dfrac{x}{M^2}\right)^\omega
   \left[
     (1+\tilde \nu) \left(\dfrac{x}{M^2}\right)^{-\omega\tilde \nu}
    -(1-\tilde \nu) \left(\dfrac{x}{M^2}\right)^{\omega\tilde \nu}
   \right], \quad
  \tilde \nu \equiv \sqrt{1-\kappa_D/\kappa_D^{\rm crit}},
\label{eq:soluc}
\end{equation}
for $\kappa_D < \kappa_D^{\rm crit}$ and
\begin{equation}
  \dfrac{B(x)}{M}
 = \frac{1}{2 i \nu} \left(\dfrac{x}{M^2}\right)^\omega
   \left[
     (1+i \nu) \left(\dfrac{x}{M^2}\right)^{-i \omega \nu}
    -(1-i \nu) \left(\dfrac{x}{M^2}\right)^{i\omega \nu}
   \right], \quad
  \nu \equiv \sqrt{\kappa_D/\kappa_D^{\rm crit}-1},
\label{eq:soloc}
\end{equation}
for $\kappa_D > \kappa_D^{\rm crit}$, 
where we find the critical $\kappa_D$
\begin{equation}
  \kappa_D^{\rm crit} \equiv \frac{1}{8} \dfrac{D-2}{D-1}.
\label{eq:critkappa}
\end{equation}
Actually, the nonoscillating solution Eq.~(\ref{eq:soluc}) for
$\kappa_D<\kappa_D^{\rm crit}$ does not satisfy the ultraviolet
boundary condition (UV-BC).
A nontrivial solution of Eq.~(\ref{eq:appsdeq}) exists only for
$\kappa_D>\kappa_D^{\rm crit}$, where the solution Eq.~(\ref{eq:soloc}) 
starts oscillating.
The critical $\kappa_D$ Eq.~(\ref{eq:critkappa}) reads
$\kappa_6^{\rm crit}=1/10$ and
$\kappa_8^{\rm crit}=3/28$, 
which are slightly smaller than the numerical results in the previous
section, Eq.~(\ref{eq:numcrit6}) and Eq.~(\ref{eq:numcrit8}).
Noting the inequality of the integral kernels 
$K_B^{\rm imp} < \tilde K_B^{\rm imp}$, however,
these results are consistent with each other.

We next turn to the scaling behavior near the critical point.
Equation (\ref{eq:soloc}) can be rewritten as
\begin{equation}
  \dfrac{B(x)}{M} 
  = \dfrac{\sqrt{1+\nu^2}}{\nu} \left(\dfrac{x}{M^2}\right)^\omega
    \sin\left[
      \theta - \omega\nu \ln \dfrac{x}{M^2}
    \right],
\qquad
    e^{i\theta} \equiv \dfrac{1+i\nu}{\sqrt{1+\nu^2}}.
\label{eq:sdsol}
\end{equation}
Inserting Eq.~(\ref{eq:sdsol}) into the UV-BC Eq.~(\ref{eq:uvbc}), we
obtain
\begin{equation}
  \theta - \omega\nu \ln\dfrac{\Lambda^2}{M^2}  +\tan^{-1}\nu =n\pi,
\label{eq:scale0}
\end{equation}
with $n$ being a positive integer.
It can be shown that the ground state corresponds to the zero-node
($n=1$) solution~\cite{MN}. 
Noting that $\theta=\tan^{-1}\nu \simeq \nu$ for $\nu\ll 1$, we thus obtain 
the scaling relation near the critical point,
\begin{equation}
  M \propto \Lambda
  \exp\left[
    \dfrac{-\pi}{(D/2-1)\sqrt{\kappa_D/\kappa_D^{\rm crit}-1}}
  \right].
\label{eq:scaling}
\end{equation}
Thus we found that the scaling of the phase transition
Eq.~(\ref{eq:scaling}) is an essential-singularity type,
the ``conformal phase transition''~\cite{MY},
similar to the result of the quenched ladder SD equation of 
four-dimensional QED~\cite{Miransky}\@. 
It is suggestive that, as we noted at the end of Section 3, 
 these SD equations are both scale invariant.

It is also worth pointing out that the essential singularity may be
used to construct models with large hierarchy between the cutoff and
the weak scale without introducing additional fine tuning.
This important property of our analysis is contrasted with the NJL
approach~\cite{ACDH}, where we need fine tuning of the NJL coupling
strength with the $(M/\Lambda)^2$ level.
(See Appendix B for details.)

Near the critical point ($\nu\rightarrow 0$ limit),
Eq.~(\ref{eq:sdsol}) gives
\begin{equation}
  B(x)
  = M \left(\dfrac{x}{M^2}\right)^{-(D/2-1)/2}
    \left( 1 +\frac{1}{2}\left(\frac{D}{2}-1\right) \ln \dfrac{x}{M^2}
    \right),
\label{eq:sdsol-crit}
\end{equation}
which is regarded as the asymptotic behavior of the solution of the
SD equation Eq.~(\ref{eq:impsdeq}).

\section{Anomalous dimension of the fermion mass}

We next consider generally the high-energy behavior of the dynamical mass when 
D$\chi$SB takes place, based on
the OPE~\cite{Pol}.

The OPE of the time-ordered fermion bilinear operator $T[\psi(x) \bar
\psi(0)]$ is given by
\begin{equation}
  -i \int d^D x e^{iq\cdot x} T [\psi^a_i(x) \bar\psi^j_b(0)]
   = c_1^M(q, g_D ; \mu) (\Gamma_M)_i{}^j \delta^a{}_b
    +c_{\bar\psi\psi}(q, g_D ; \mu) \delta_i{}^j \delta^a{}_b 
     (\bar\psi\psi)
    +\cdots, \label{OPE}
\end{equation}
with $c_1$, $c_{\bar\psi\psi}$ being the Wilson coefficient functions,
where $a,b$ are for gauge indices 
and $i,j$ are for spinor indices. 

It is straightforward to evaluate the Wilson coefficient function
$c_{\bar\psi\psi}$ in the $D$-[$=(4+\delta)$-]-dimensional gauge theories
at  the tree level, 
\begin{equation}
  c_{\bar\psi\psi}(q, \hat g; \mu)
  = \dfrac{(D-1)}{\eta}\dfrac{C_F}{N}
    \dfrac{\hat g^2}{\mu^\delta} \dfrac{1}{q^4},
\end{equation}
where we adopted the Landau gauge.
Comparing  Eq.~(\ref{OPE}) with the propagator of the fermion field
\begin{equation}
 -iS(p) = \frac{1}{A(-p^2)\fsl{p}-B(-p^2)} 
  \simeq \frac{\fsl{p}}{A(-p^2)p^2}+\frac{B(-p^2)}{A^2(-p^2)p^2}
  + \cdots,
\end{equation}
we find that the high-energy behavior of the dynamical fermion mass function
$B(-p^2)$ is given by 
\begin{equation}
  B(-p^2) \simeq p^2 c_{\bar\psi\psi}(p, \hat g; \mu) 
  \langle \bar{\psi} \psi \rangle,
  \label{massfn}
\end{equation}
where we have assumed  absence of wave-function renormalization 
of the fermion field, which is justified in the Landau gauge within the
ladder approximation. 

The solution of 
the RGE for
$c_{\bar\psi\psi}$ is given by,
\begin{equation}
  \left[
    \dfrac{\partial}{\partial t}
   -\hat \beta \dfrac{\partial}{\partial \hat g}
   + D - \gamma_m(\hat g)
  \right] c_{\bar\psi\psi}(e^t p, \hat g; \mu)  =0,
\label{eq:rge3} 
\end{equation}
which is solved as 
\begin{equation}
  c_{\bar\psi\psi}(e^t p, \hat g; \mu)
  = c_{\bar\psi\psi}(p, \bar g(t); \mu) 
    \exp\int_0^t dt \left[\gamma_m(\bar g(t)) - D\right],
\end{equation}
with the running gauge coupling $\bar g(t)$:
\begin{equation}
  \bar g^2(t) 
   = \dfrac{1}{
       \dfrac{e^{-\delta t}}{\hat g^2(\mu)}
      -\left(\dfrac{2}{\delta}+1\right) \Omega_{\rm NDA} b
        \left[ 1 - e^{-\delta  t}\right]
     }.
\end{equation}
For sufficiently large $t$, it is evident that
$\lim_{t\rightarrow\infty} \bar g(t) = g_*$.
The high-energy behavior of the Wilson coefficient function
$c_{\bar\psi\psi}$ therefore reads 
\begin{equation}
  c_{\bar\psi\psi}(e^t p, \hat g; \mu)
  \propto e^{(\gamma_m^* - D)t},
  \qquad
  \gamma_m^* \equiv \gamma_m(g_*),
\end{equation}
or
\begin{equation}
  c_{\bar\psi\psi}(p, \hat g; \mu)
  \propto (-p^2)^{(\gamma_m^* - D)/2}.
\end{equation}
The high-energy behavior of the mass function $B$ in Eq.~(\ref{massfn})
thus is  given by
\begin{equation}
  B(-p^2) \propto (-p^2)^{(\gamma_m^* + 2 - D)/2}.
\label{eq:masspow}
\end{equation}

\begin{figure}[t]
  \begin{center}
    \includegraphics{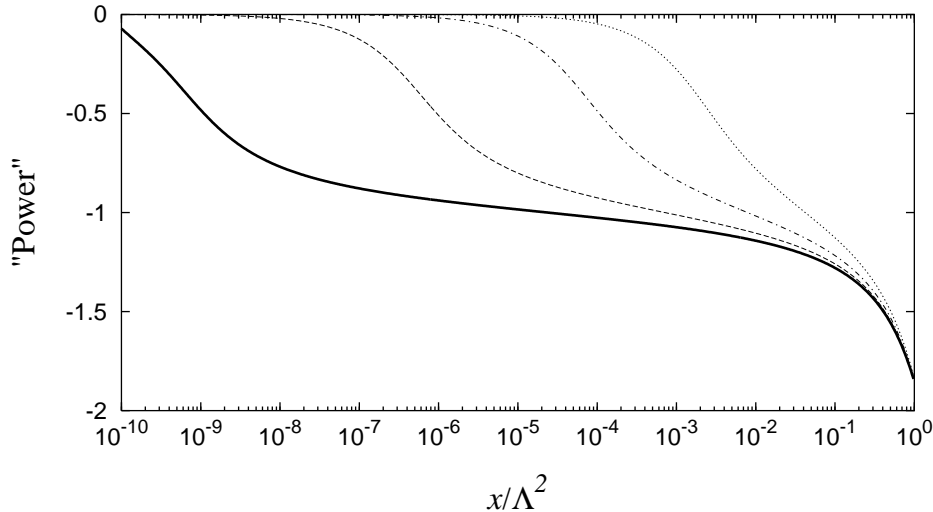}
    \caption{
      The ``power'' behavior of the mass function in six dimensions. 
      The lines from left to right represent graphs for
      $\kappa_6=0.122, 0.125, 0.130,0.140$ 
      [or $B^2(M_0^2)/\Lambda^2=4.1\times 10^{-10},5.4\times 10^{-7},
      6.9\times 10^{-5},2.2\times 10^{-3}$] with $\Lambda^2/M_0^2=10^{10}$. 
      \label{plot_gamma_6dim}
    }
  \end{center}
\end{figure}

The anomalous dimension at the fixed point $\gamma_m^*$ can be
extracted from the numerical solution of the SD equation.
For this purpose we define the ``power'' of the mass function
($\omega$)
\begin{equation}
  \omega \equiv \frac{x}{B(x)}\frac{d}{dx} B(x).
\end{equation}

Figure \ref{plot_gamma_6dim} shows the ``power'' behaviors of the
numerical solution of the SD equation in six dimensions
for various ``couplings.''
It can be seen that the ``power'' is almost constant in the asymptotic 
region $B^2(M_0) \ll x \ll \Lambda^2$ as we expected from
Eq.~(\ref{eq:masspow}).
The behavior near the cutoff $x\simeq \Lambda^2$ in Fig. 4
is an artifact~\cite{NT} 
due to the sharp
cutoff introduced in the analysis of the SD equation.\footnote{
Equation (\ref{eq:uvbc}) leads to the relation of 
$\Lambda^2 B'(\Lambda^2)/B(\Lambda^2)=1-D/2$. } 
This artifact disappears at the limit of $\Lambda \to \infty$. 
Reading the ``power'' in the asymptotic region ($\omega\simeq -1$), we
obtain
\begin{equation}
  \gamma_m^* \simeq 2\omega + (D-2) \simeq 2
\end{equation}
for the $D=6$ bulk gauge theory at the critical point $\kappa_6^{\rm crit}$.

\begin{figure}[t]
  \begin{center}
    \includegraphics{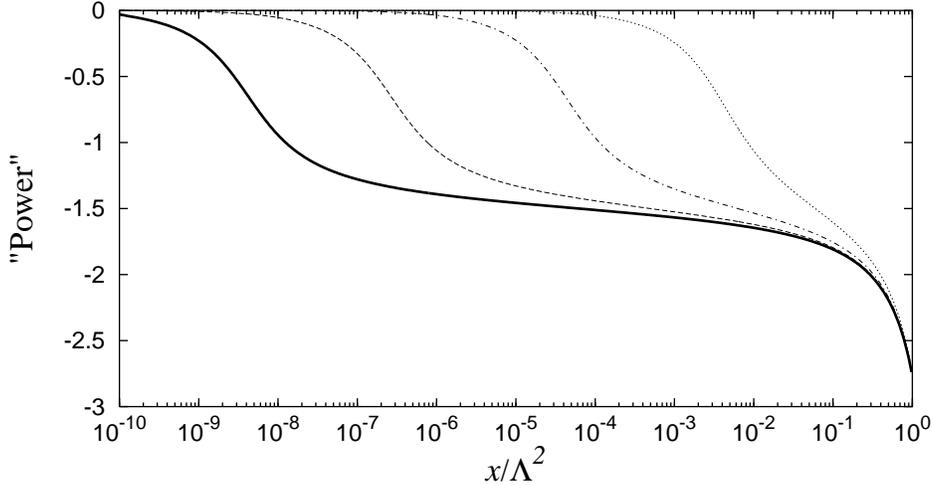}
    \caption{
      The ``power'' behavior of the mass function in eight dimensions. 
      The lines from left to right represent graphs for
      $\kappa_8=0.146, 0.147, 0.150, 0.160$ 
      [or $B^2(M_0^2)/\Lambda^2=5.8\times 10^{-9},4.1\times 10^{-7},
      6.8\times 10^{-5},6.8\times 10^{-3}$]
      with $\Lambda^2/M_0^2=10^{10}$. 
      \label{plot_gamma_8dim}
    }
  \end{center}
\end{figure}

A similar analysis is also performed for $D=8$.
The corresponding ``power'' behavior is shown in
Fig.~\ref{plot_gamma_8dim}.
The anomalous dimension is then
\begin{equation}
  \gamma_m^* \simeq 2\omega + (D-2) \simeq 3,
\end{equation}
for $D=8$.

The analytical result in the previous section
Eq.~(\ref{eq:sdsol-crit})
compared with Eq.~(\ref{eq:masspow}) yields 
\begin{equation}
  \gamma_m^*=\frac{D}{2}-1, 
\label{eq:gamma_m}
\end{equation}
which agrees with 
the above numerical result.

It is remarkable that 
Eq.~(\ref{eq:gamma_m}) is also consistent with the conformal phase
transitions for other dimensions
$D\leq 4$:
$\gamma_m=1/2$ for $D=3$ agrees with the high-energy behavior~\cite{ANW}
and $\gamma_m =1$ for $D=4$ is the walking theory~\cite{Hol} obtained from the
ladder SD equation with fixed coupling.
They are obtained in different approximations: Namely, 
the result for three dimensions is obtained by running coupling  
with the IR fixed point and  that for 
six/eight dimensions by running coupling with the UV fixed point,
while the four-dimensional result is obtained by fixed coupling.
However, the SD equations in all these cases happen to be quite similar 
because of the {\it scale invariance at the fixed point}. 

It should be emphasized that
such a  large anomalous dimension implies suppression of the 
flavor-changing-neutral-current problem
in the dynamical electroweak symmetry breaking scenario as in 
walking technicolor~\cite{Hol}.
The large $\gamma_m$ observed in this section is, therefore, good
news for constructing phenomenologically viable models in this direction.

Moreover, the corresponding
asymptotic behavior of the mass function Eq.~(\ref{eq:sdsol-crit}), 
$B(p_E^2) \sim M^{D/2} (p_E^2)^{(1-D/2)/2}$,
still yields strong convergence of the bulk decay constant $F^{(D)}_\pi$, 
which may be calculated through the 
PS formula~\cite{PS}:
\begin{eqnarray}
(F^{(D)}_\pi)^2 
  &\sim& \int\dfrac{d^D p_E}{(2\pi)^D}
             \dfrac{B^2(p_E^2)}{[p_E^2+B^2(p_E^2)]^2}, 
  \nonumber\\
  &\propto& M^D \int dp_E \frac{1}{p_E^3}.  
\label{eq:PSD}
\end{eqnarray}
This suggests that the dynamically induced bulk Yukawa coupling 
$g_Y=M/F^{(D)}_\pi$
can be made finite even in the ``infinite cutoff limit'' $\Lambda \rightarrow 
\infty$.\footnote{This statement is of course rather formal  
in the sense that $\kappa_D$ in Eq.~(\ref{eq:kappa_def}) is
actually not an arbitrary adjustable parameter and hence cannot be fine tuned to 
$\kappa_D^{\rm crit}$, $\kappa_D \rightarrow \kappa_D^{\rm crit}$,
to make the dynamical mass $M$ finite through Eq.~(\ref{eq:scaling})
in that limit.
}
This is in contrast with
perturbation theory where the gauge theory in $D \;(>4)$ dimensions 
is obviously nonrenormalizable.
This situation is similar to the renormalizability of the gauged NJL 
model~\cite{KTY}.

\section{Chiral fermion on the 3-brane}

We have investigated so far the possibility of D$\chi$SB in the bulk.
Since a chiral fermion in the bulk ($D>4$) has four or 
more components, 
it is nontrivial to obtain a four-dimensional chiral fermion with two 
components as an effective theory.
For such a purpose, we need to compactify the extra dimensions on an
orbifold, in which unwanted components are projected out by its
boundary conditions~\cite{ACDH}.
In this section we describe a systematic procedure to find such
orbifold compactifications.

We start with the minimal case $D=6$ for simplicity.
The chiral projection operators in six dimensions are given by
\begin{equation}
  \dfrac{1\pm \Gamma_{A,7}}{2}, \qquad
  \Gamma_{A,7} \equiv \Gamma^0 \Gamma^1 \Gamma^2 \Gamma^3
                      \Gamma^5 \Gamma^6,
\end{equation}
and the chiral fermions $\psi_{\pm}$ obey
\begin{equation}
  \Gamma_{A, 7} \psi_{\pm} = \pm \psi_{\pm}.
\end{equation}
Hereafter we argue only $\psi_+$, the chiral fermion with positive
chirality in the bulk.
It is easy to extend our arguments to the case of $\psi_-$.

We next decompose the space-time coordinate into conventional and
extra dimensions:
\begin{equation}
  x^M = (x^\mu, y^m), \qquad
  \mu = 0,1,2,3, \qquad
  m = 5,6,
\end{equation}
and assume a torus compactification,
\begin{equation}
  \psi_+(x,y^5, y^6) = \psi_+(x,y^5+2\pi R, y^6) 
  = \psi_+(x,y^5, y^6+2\pi R),
\end{equation}
where the radii of the fifth and sixth dimensions are assumed to be the
same (denoted by $R$) for simplicity.
The chiral fermion in $D=6$ is then decomposed into KK modes:
\begin{equation}
  \psi_+(x,y) = \sum_{k_5,k_6} \psi^{k_5 k_6}_+(x) 
    \exp\left[i \dfrac{k_5 y^5 + k_6 y^6}{R} \right].
\end{equation}

We next introduce the four-dimensional chirality matrix $\Gamma_{A,5}
\equiv i\Gamma^0 \Gamma^1 \Gamma^2 \Gamma^3$.
It is easy to show several identities:
\begin{equation}
  [\Gamma_{A,5}, \Gamma_{A,7}] = 0, \quad
  \Gamma_{A,5} \Gamma_{A,5} = 1, \quad
  \tr\left[\Gamma_{A,5}\dfrac{1 \pm \Gamma_{A,7}}{2}\right] = 0,
\end{equation}
which indicate that $\Gamma_{A,5}$ and $\Gamma_{A,7}$ are
simultaneously diagonalizable, eigenvalues of $\Gamma_{A,5}$ are $\pm
1$, and the sum of eigenvalues of $\Gamma_{A,5}$ is zero for a
chiral fermion in the six-dimensional bulk.
It is therefore evident that the zero mode $\psi_+^{00}$ in this
torus compactification  is  vectorlike in its four-dimensional
effective theory.
We need to eliminate unwanted components of the fermion on the
four-dimensional brane by imposing a certain orbifold symmetry.

It should be noted, however, that the ``parity'' of extra dimensions
does not suit our purpose, because it is explicitly violated in the
chiral theory of the bulk.\footnote{
$CP$ is also violated in the six-dimensional bulk, since charge
conjugation does not flip chirality in $D=4k+2$ dimensions.}
We then try to adopt rotation in the extra dimensions by the angle
$\pi$:\footnote{
It is also possible to use $\pi/2$ rotation to define an orbifold,
which keeps chirality on the four-dimensional brane.}
\begin{eqnarray}
  \psi'(x,y^5,y^6)
  &=& \exp\left[\frac{i}{2} \Sigma^{56}\pi \right]\psi(x,-y^5,-y^6)
  \nonumber\\
  &=& i\Sigma^{56} \psi(x, -y^5, -y^6),
\end{eqnarray}
with $\Sigma^{MN}$ being defined by 
\begin{equation}
  \Sigma^{MN} \equiv \frac{i}{2} [ \Gamma^M, \Gamma^N ].
\end{equation}

There are two possible boundary conditions of this orbifold:
\begin{equation}
  \psi_+(x, y^5, y^6) = (-1)^n \Sigma^{56} \psi_+(x, -y^5, -y^6),
  \qquad
  \mbox{$n=0$ or $1$},
\label{eq:boundary}
\end{equation}
which leads to the constraint for the zero mode fermion
\begin{equation}
  \psi_+^{00}(x) = (-1)^n \Sigma^{56} \psi_+^{00}(x).
\label{eq:4chiral}
\end{equation}
Noting the identity $\Sigma^{56} = - \Gamma_{A,5} \Gamma_{A,7}$,
we can rewrite Eq.~(\ref{eq:4chiral}) into the conditions of the chiral 
fermion on the four dimensional brane:
\begin{equation}
  \Gamma_{A,5} \psi_+^{00}(x) = (-1)^{n+1} \psi_+^{00}(x).
\end{equation}
The chirality on the brane is determined by the choice of
the boundary condition, $n=0$ or $1$, in Eq.~(\ref{eq:boundary})  

It should be emphasized here that our procedures described in this
section do not depend on a particular choice of the representation of
the Clifford algebra. 
We can easily generalize our arguments to an orbifold
compactification from $D=2(k+1)$ into $D=2k$ dimensions.
By applying these procedures repeatedly, we are thus able to obtain
orbifold compactification starting from a bulk chiral theory of $D=2k$
$(k\ge 3)$ into a brane chiral theory with four dimensions.

\section{Effective gauge coupling }

Although the $\overline {MS}$-scheme has been widely
adopted for running gauge coupling  in the improved
ladder approximation, it is worth investigating yet another
choice, ``effective'' gauge coupling $g_{\rm eff}$, which is closely
related to the gauge boson propagator and its momentum.
For this purpose, we evaluate the one-loop gauge boson propagator in
the truncated KK effective theory on the 3-brane, and derive a
relation between the effective and the $\overline{MS}$ couplings. 

The effective gauge coupling  $g_{\rm eff}$ on the 3-brane is defined
by\footnote{
We use the background gauge-fixing method throughout this
section. 
The Ward-Takahashi identities of non-Abelian gauge theory are QED-like
and keep manifest gauge invariance in this gauge fixing.}
\begin{equation}
  \dfrac{-i}{g_{\rm eff}^2(q^2)} D_{\mu\nu}^{-1}(q)
  \equiv 
  \dfrac{-i}{g_0^2} D_{(0)\mu\nu}^{-1}(q)
  - ( q^2 g_{\mu\nu} - q_\mu q_\nu )\Pi(q^2)
\label{eq:eff1}
\end{equation}
with $g_0$ being the bare gauge coupling, and the (four-dimensional) gauge
boson propagators are given by
\begin{equation}
  D_{(0)\mu\nu}(q) = \dfrac{-i}{q^2}\left(
    g_{\mu\nu} - (1 - \xi_0) \dfrac{q_\mu q_\nu}{q^2}
  \right), \qquad
  D_{\mu\nu}(q) = \dfrac{-i}{q^2}\left(
    g_{\mu\nu} - (1 - \xi_{\rm eff}(q^2)) \dfrac{q_\mu q_\nu}{q^2}
  \right).
\end{equation}
Equation (\ref{eq:eff1}) reads
\begin{equation}
  \dfrac{1}{g_{\rm eff}^2(q^2)}
  = \dfrac{1}{g_0^2} - \Pi(q^2).
\label{eq:eff2}
\end{equation}
The vacuum polarization function $\Pi(q^2)$ can be decomposed into 
loops of each KK mode at the one-loop level:
\begin{equation}
  \Pi(q^2) = \sum_{\vec n} \Pi(q^2, m_{\vec n}^2), \qquad
  m_{\vec n}^2 = \dfrac{|\vec n|^2}{R^2}.
\end{equation}

In order to calculate the relation between the effective and the
$\overline{MS}$ couplings, we next evaluate $\Pi(q^2, m_{\vec n}^2)$
using dimensional regularization ($d\equiv 4+\epsilon$),
\begin{equation}
  \Pi( q^2, m^2 ) = C_G \left[ 
    4I_g(q^2, m^2) + (2-D)I_b(q^2, m^2)
  \right] - 2\eta T_R N_f I_f(q^2, m^2),
\end{equation}
where we used notations introduced in Section 2 and
\begin{eqnarray*}
  I_g(q^2, m^2) 
    &\equiv& \dfrac{\Gamma(2-d/2)}{(4\pi)^{d/2}}
             \int_0^1 dx \left[ m^2 - x(1-x) q^2 \right]^{d/2-2}, \\
  I_b(q^2, m^2) 
    &\equiv& \dfrac{\Gamma(2-d/2)}{2(4\pi)^{d/2}}
             \int_0^1 dx (2x-1)^2 \left[ m^2 - x(1-x) q^2
    \right]^{d/2-2}, \\
  I_f(q^2, m^2)
    &\equiv& \dfrac{\Gamma(2-d/2)}{(4\pi)^{d/2}}
             \int_0^1 dx x(1-x) \left[ m^2 - x(1-x) q^2 \right]^{d/2-2}.
\end{eqnarray*}
The counterterm for the $\overline{MS}$ coupling\footnote{
Strictly speaking, Eq.~(\ref{eq:msbar}) is evaluated in the modified
dimensional reduction scheme~\cite{kn:DRbar}.
} 
in the truncated KK effective theory is given by
\begin{equation}
  \dfrac{1}{g_{\overline{MS}}^2 (\mu)} =
    \dfrac{1}{g_0^2}
    - \sum_{\vec n}^{m_{\vec n}>\mu} \Pi(q^2=0, m_{\vec n}^2)
    + \sum_{\vec n}^{m_{\vec n}\leq \mu} 
      \dfrac{\Gamma(2-d/2)}{(4\pi)^{d/2}} b' \mu^{d-4},
\label{eq:msbar}
\end{equation}
where the term 
$\sum_{\vec n}^{m_{\vec n}>\mu} \Pi(q^2=0, m_{\vec n}^2)$ comes from
the loop of KK modes heavier than the renormalization scale $\mu$.
This term is independent of $\mu$ and therefore does not affect
the RGE for the gauge coupling,
in accordance with the decoupling theorem that is assumed in the
truncated KK effective theory.

Taking the $d\rightarrow 4$ limit we now obtain 
\begin{equation}
  \dfrac{1}{g_{\rm eff}^2(q^2)}
  = \dfrac{1}{g_{\overline{MS}}^2(\mu)}
    - \sum_{\vec n}^{m_{\vec n}>\mu} \Pi_{>}(q^2, m_{\vec n}^2)
    - \sum_{\vec n}^{m_{\vec n}\leq \mu} \Pi_{<}(q^2, m_{\vec n}^2;
  \mu),
\label{eq:eff}
\end{equation}
where $\Pi_{>}$ and $\Pi_{<}$ are given by
\begin{eqnarray}
  (4\pi)^2 \Pi_{>}(q^2, m^2)
  &\equiv & -C_G \int_0^1 dx \left[ 4 + \frac{2-D}{2}(2x-1)^2 \right]
       \ln\left(1-\dfrac{q^2}{m^2}x(1-x)\right) \nonumber\\
  & &+ 2\eta T_R N_f \int_0^1 dx x(1-x) 
       \ln\left(1-\dfrac{q^2}{m^2}x(1-x)\right), 
  \\
  (4\pi)^2 \Pi_{<}(q^2, m^2; \mu)
  &\equiv & -C_G \int_0^1 dx \left[ 4 + \frac{2-D}{2}(2x-1)^2\right]
       \ln\left(\dfrac{m^2 - q^2 x(1-x)}{\mu^2} \right)
  \nonumber\\
  & &+2\eta T_R N_f \int_0^1 dx x(1-x) 
       \ln\left(\dfrac{m^2 - q^2 x(1-x)}{\mu^2} \right).
\end{eqnarray}
Note here that the $\mu$ dependence in the right-hand side of Eq.~(\ref{eq:eff})
cancels exactly. 
We also introduce a {\it dimensionless} effective gauge coupling of the
{\it bulk} gauge theory defined in a similar manner to
Eq.~(\ref{eq:dimless}),
\begin{equation}
  \hat g^2_{\rm eff}(q^2) 
    = \dfrac{(2\pi R\sqrt{-q^2})^\delta}{n} g^2_{\rm eff}(q^2).
\label{eq:effdimless}
\end{equation}

Using approximations described in Appendix \ref{sec:app2}, we find
\begin{equation}
  \dfrac{1}{\hat g_{\rm eff}^2(q^2)} \simeq
    \dfrac{\lambda}{\hat g_{\overline{MS}}^2(\sqrt{-\lambda q^2})}
   +\dfrac{1}{(4\pi)^3} 
    \left[-\frac{3}{5} C_G +\frac{\eta}{15} T_R N_f \right]
    \ln \left(\dfrac{-\lambda q^2}{\Lambda^2}\right)
\label{eq:sixdimeff}
\end{equation}
for the bulk gauge theory in $D=6$ dimensions.
Here $\lambda$ is given by
\begin{equation}
  \lambda 
   = \exp\left[
       \dfrac{C_G\left( -8 -\frac{4}{9} (2-D)\right)
                +\frac{5}{9} \eta T_R N_f}{-b'}
     \right].
\label{eq:optla2}
\end{equation}
\begin{figure}[t]
  \begin{center}
    \includegraphics{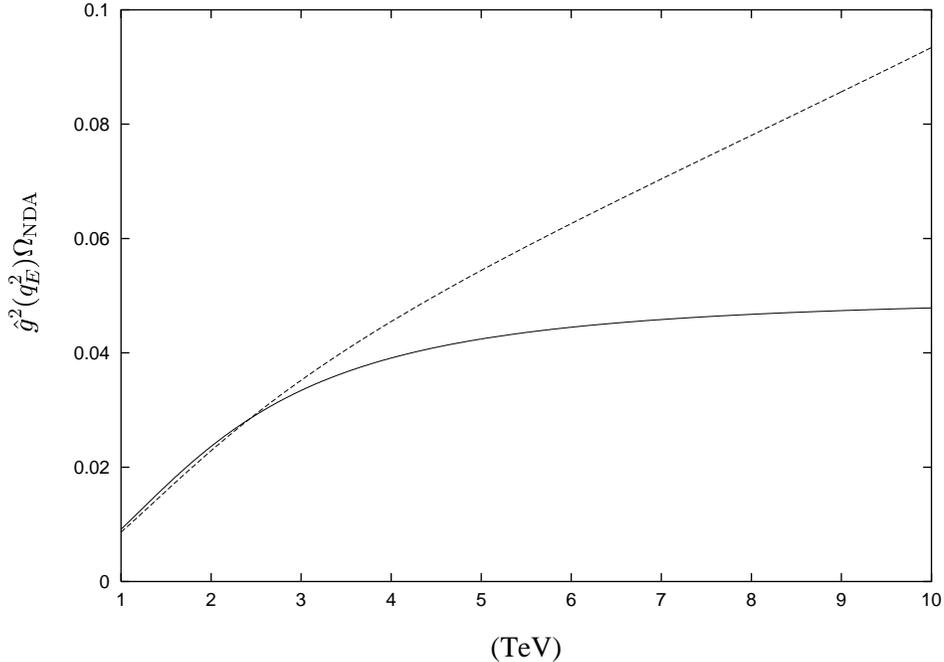}
    \caption{
      The graph of the dimensionless gauge coupling
      with  $C_G=3$, $N_f=0$, $R^{-1}=1$ TeV, $\Lambda=10$ TeV,
      $\alpha_{\overline{MS}}(M_Z)=0.1$. 
      The solid line and the dashed line represent the $\overline{MS}$
      coupling and the effective coupling of Eq.~(\ref{eq:sixdimeff}), 
      respectively. 
    \label{fig:plot_alpha}}
  \end{center}
\end{figure}
In the analysis of the SD equation based on $g_{\rm eff}$, we
adopt the effective coupling $\hat g^2_{\rm eff}$
Eq.~(\ref{eq:sixdimeff}) [instead of $\hat g^2_{\overline{MS}}$ in
Eq.~(\ref{ghat})] in the formula for the improved ladder
approximation Eq.~(\ref{g_D}). 

Several comments are in order.

a) The decoupling theorem is violated in the effective coupling
Eq.~(\ref{eq:sixdimeff}), since it depends explicitly on the
ultraviolet cutoff $\Lambda$.
This result comes from the nonrenormalizability of 
the six-dimensional bulk gauge theory. 

b) The effective coupling  is larger than the $\overline{MS}$ coupling by
approximately a factor of $\lambda^{-1}$. (See also 
Fig.~\ref{fig:plot_alpha}.) 
If we adopt $\hat g_{\rm eff}$ instead of $\hat g_{\overline{MS}}$ in
the improved ladder approximation, there is a chance that the
bulk QCD coupling even in six dimensions 
can be strong enough to cause D$\chi$SB in the bulk
under certain conditions.

c) There still exists an {\it upper bound} on $\hat g_{\rm eff}$
similar to the nontrivial UV fixed point, which is roughly
proportional to the UV fixed point in the $\overline{MS}$ scheme.
(See Appendix \ref{sec:app2} for a detailed discussion.) 
It is therefore still a nontrivial question whether D$\chi$SB occurs
or not in the bulk gauge theories even if we adopt $\hat g_{\rm eff}$
in the improved ladder SD equation.

d) We can define an analogue of the ``$\beta$ function'' for ``bare coupling'' 
$\hat g_\Lambda \equiv \hat g_{\rm eff}(q^2 = -\Lambda^2)$. (See Appendix
\ref{sec:app2}.)
The upper bound of $\hat g_{\rm eff}$ can be regarded as an UV fixed point 
of such a ``$\beta$ function'' and therefore independent of the choice 
of the cutoff scale $\Lambda$.

In fact the finite renormalization effect is the largest uncertainty
of our analysis based on the improved ladder SD equation, compared with
other uncertainties
such as the nonladder effects, higher-order corrections, etc.
A detailed analysis of the improved ladder SD equation with 
$\hat g_{\rm eff}$ will be presented elsewhere~\cite{HTY}.

\section{Summary and discussions}

We have studied dynamical issues of the ACDH version~\cite{ACDH} of the
TMSM~\cite{MTYa,MTYb,nambu,BHL} 
within the framework of the improved
ladder SD equation. 
Based on the 
truncated KK effective theory~\cite{DDG}, we found 
that $D$-dimensional non-Abelian gauge theories
with compactified extra dimensions possess  a nontrivial UV fixed point.  
We then evaluated the UV fixed point by using the one-loop RGE, assuming
its nonperturbative existence.
Although the SM couplings in the $D$-dimensional
bulk generally become strong beyond the compactification scale, 
the (dimensionless) bulk coupling cannot grow beyond the UV fixed point
and hence it is highly nontrivial whether or not D$\chi$SB really takes place.
 
For the simplest scenario of ACDH with the 
massless gauge bosons and third family
quarks and leptons living in the $D$-dimensional bulk and the rest in the
four-dimensional brane (3-brane), we have the UV fixed points 
Eqs.~(\ref{eq:ACDH6}),(\ref{eq:ACDH8}):
\begin{equation}
\kappa_6=0.091, \quad \kappa_8=0.242
\end{equation}
for bulk dimensions of six and eight, respectively.

On the other hand, the improved ladder SD equation in $D \;(>4)$ dimensions
yields  the critical points in six/eight dimensions as 
Eqs.~(\ref{eq:numcrit6}),(\ref{eq:numcrit8}): 
\begin{equation}
\kappa_6^{\rm crit}=0.122, \quad \kappa_8^{\rm crit}=0.146.
\end{equation}
These results are qualitatively consistent with a naive dimensional
analysis. 
The ACDH scenario thus can work for $D=8$ but not for $D=6$ if we take the 
results Eq.~(8.1) and Eq.~(8.2) at face value.
It should be emphasized, however, that our analysis is based on the
ladder approximation. 
We would certainly need further investigation to incorporate non-ladder
effects in order to evaluate the critical value more accurately. 

We also discussed some subtlety about the ``improved'' ladder SD equation
by replacing the $\overline {MS}$ running coupling by the effective coupling
including the finite renormalization effects.
This makes attractive forces somewhat larger
than in $\overline {MS}$ coupling, so that the condensate 
can occur more easily. In the case of the ACDH scenario,
top condensation may be possible due to effects of 
finite renormalization even in six dimensions, since the coupling has
a chance to increase over the critical point {\it for sufficiently large
cutoff}.

However, 
if the cutoff is too large, then the  $U(1)$ coupling 
dominates the QCD coupling so 
that the MAC favors other channels (tau lepton condensate) 
than the top quark condensate.
We then obtain some conditions for the ``effective (phenomenological) 
cutoff'' ($\Lambda$) where the bulk QCD and hypercharge couplings are
aligned in the MAC in such a way that only the top quark condenses while others 
(bottom, etc.) do not. 
Another constraint comes from the top quark mass which is related to
the decay constant $F_\pi^{(D)}$ in $D$ dimensions through the PS formula, 
Eq.~(\ref{eq:PSD}). Thus it is related to the weak scale $F_\pi=246 \;{\rm GeV}$ 
as
\begin{equation}
F^2_\pi = \frac{(2\pi R)^{\delta}}{n} (F^{(D)}_\pi)^2.
\end{equation}
These matters will be dealt with in a forthcoming paper~\cite{HTY}.

The salient feature of the improved ladder SD equation in $D\;(>4)$
dimensions is its approximate scale invariance. 
The reason is the followings. The bulk
dimensionful coupling can be written as the dimensionless coupling
multiplied by a factor having dimensions carried by the 
renormalization point $\mu$ [see Eq.~(\ref{eq:dimless2})], which is then
traded 
for the momentum in the running coupling in the improved ladder SD 
equation, and moreover the running coupling quickly increases up to
the UV fixed point. Hence the SD equation can be well approximated by the 
coupling at the fixed point. Then the resulting SD equation
Eq.~(\ref{eq:impsdeq})
has no scale parameters  except for the cutoff.

This is the very reason why we obtained the essential-singularity scaling
of the conformal phase transition, Eq.~(\ref{eq:scaling}), 
with the analytical result for the critical
point Eq.~(\ref{eq:critkappa})
\begin{equation}
\kappa_D^{\rm crit}=\frac{D-2}{8(D-1)},
\end{equation}
which was
derived through 
certain approximations and 
is consistent with the numerical result above.

The essential-singularity scaling gives us the possibility to
have a large hierarchy between the weak scale and the cutoff
without fine tuning. Here we note that $\kappa_D$ is not an arbitrary
parameter but a definite number once the model is set up.
We should therefore note that it is impossible to take the cutoff
infinitely large. 

In a realistic model based on the ACDH scenario, however,
the bulk gauge coupling of QCD is determined through matching with
the QCD on the 3-brane at the compactification scale $R^{-1}$.
The bulk coupling grows at high energy toward the UV fixed point
and can exceed the critical coupling only for a certain cutoff
(``critical cutoff'').
When we tune the cutoff very close 
to the critical one, the SD equation yields a very small dynamical mass
compared with the cutoff.
We are thus able to determine the value of the cutoff, which enables
us to evaluate the low-energy predictions of the ACDH scenario (e.g.,
$m_t$ and $m_H$) more accurately. 

Moreover, we had a very large anomalous dimension Eq.~(\ref{eq:gamma_m}):
\begin{equation}
 \gamma_m = \frac{D}{2}-1, 
\end{equation}
which happens to coincide with the cases for $D\leq 4$, i.e.,
the quenched ladder SD equation (with
nonrunning or walking/standing coupling) for $D=4$ 
($\gamma_m=1$) and
also with the improved ladder SD equation for $D=3$ QED ($\gamma_m=1/2)$
where the running coupling has an infrared fixed point. In all the cases
including $D\leq 4$, the SD equation has scale invariance
at the fixed point. 

\section*{Acknowledgments}
We would like to thank Howard Georgi, Valery Gusynin, Masayasu Harada,
Yoshio Kikukawa, Volodya Miransky, Takeo Moroi, and Masahiro Yamaguchi for
useful discussions. 
K.Y. thanks Howard Georgi for hospitality at Harvard where a part of
this work was done. This work is supported by Grant-in-Aid for
Scientific Research (B) $\#$11695030 (K.Y.), (A) $\#$12014206 (K.Y.),
the JSPS Research Fellowships for Young Scientists $\#$01170 (M.H.), 
and the Oversea Research Scholar Program of the Monbusho
(Ministry of Education, Science, Sports and Culture) (K.Y.).

\appendix
\section{Angular integrals in the ladder SD equations}
\label{sec:app1}

The momentum integrals of the SD equations
Eq.~(\ref{eq:sda}) and Eq.~(\ref{eq:sdb}) can be decomposed into polar
and angular integrals,
\begin{equation}
 \int \dfrac{d^D q_E}{(2\pi)^D}F(p_E^2,q_E^2,p_E \cdot q_E)
  = C_D \int_0^{\Lambda^2} dy\, y^{D/2-1}
    \int_{0}^{\pi} d\theta \sin^{D-2}\theta 
    F(x,y,\sqrt{xy}\cos\theta) ,
\end{equation}
with $x \equiv p_E^2$, $y \equiv q_E^2$, and $C_D$ defined by
$C_D \equiv \Omega_{\rm NDA}/B(1/2,D/2-1/2)$. 
In order to evaluate the angular integral $d\theta$, we define the
integral
\begin{equation}
 I(\mu,\nu,\rho; z) \equiv \int_{0}^{\pi} d\theta 
 \frac{\sin^{2\nu+1}\theta \cos^{\rho}\theta}
      {\left(z-\cos\theta\right)^{\mu+\nu+1}}.
\end{equation}
It is easy to see that the angular integrals in Eq.~(\ref{eq:sda}) and
Eq.~(\ref{eq:sdb}) can be expressed in terms of $I(\mu,\nu, 0; z)$ and
$I(\mu,\nu,1; z)$ with $z\equiv (x+y)/(2\sqrt{xy})$.
We obtain~\cite{kondo}
\begin{eqnarray}
  A(x) &=& 1 + \dfrac{C_F g_D^2}{x} C_D \int_0^{\Lambda^2}
    dy y^{D/2-1} \dfrac{A(y)}{A^2 y + B^2} 
       \nonumber\\
  & & \times \left\{
    \frac{D-1-\xi}{2} 
       I(\frac{3}{2}-\frac{D}{2}, \frac{D}{2}-\frac{3}{2}, 1; z)
   -\frac{1-\xi}{2}
       I(\frac{3}{2}-\frac{D}{2}, \frac{D}{2}-\frac{1}{2}, 0; z)
    \right\},
\label{eq:appsda1}
  \\
  B(x) &=& (D-1+\xi) C_F g_D^2 C_D \int_0^{\Lambda^2}
    dy y^{D/2-1} \dfrac{B(y)}{A^2 y + B^2} 
    \dfrac{1}{2\sqrt{xy}} 
    I(\frac{3}{2}-\frac{D}{2}, \frac{D}{2}-\frac{3}{2}, 0; z).
  \nonumber\\
  & &
\label{eq:appsdb1}
\end{eqnarray}
Using the relation
\begin{equation}
 I(\mu,\nu, 1; z) =
   \dfrac{1}{2\nu+2} \int_{0}^{\pi} d \theta 
   \dfrac{1}{\left(z-\cos\theta\right)^{\mu+\nu+1}}
   \frac{d}{d\theta} \left[ \sin^{2\nu+2} \theta \right]
  = \dfrac{\mu+\nu+1}{2\nu+2} I(\mu,\nu+1,0;z),
\end{equation}
Eq.~(\ref{eq:appsda1}) can be further simplified:
\begin{equation}
  A(x) = 1 + \dfrac{\xi}{2} \dfrac{D-2}{D-1} 
             \dfrac{C_F g_D^2}{x} C_D \int_0^{\Lambda^2}
             dy y^{D/2-1} \dfrac{A(y)}{A^2 y + B^2} 
             I(\frac{3}{2}-\frac{D}{2}, \frac{D}{2}-\frac{1}{2}, 0; z).
\end{equation}
It should be noted that $A(x)=1$ holds for arbitrary dimensions in the
Landau gauge $\xi=0$~\cite{kondo}.

The integral $I(\mu,\nu,0; z)$ was given for certain integer dimensions
in Ref.~\cite{kondo} and now is  expressed in terms of the
hypergeometric function $F(\alpha,\beta,\gamma; z)$ for arbitrary $D$:
\begin{equation}
 I(\mu,\nu,0 ;z) 
 =  
 \dfrac{2^{\mu+\nu+1}\sqrt{\pi}\,\Gamma(\nu+1)}
      {\tilde{z}^{\mu+\nu+1} \Gamma(\nu+\frac{3}{2})}  
  \hyper{\mu+\nu+1}{\mu+\frac{1}{2}}{\nu+\frac{3}{2}}{\tilde{z}^{-2}},
\label{Ia1}
\end{equation}
with 
\begin{equation}
  \tilde z \equiv z + \sqrt{z^2-1} = \dfrac{\max(x,y)}{\sqrt{xy}},
  \qquad
   \tilde z^{-2} = \dfrac{\min(x,y)}{\max(x,y)}.
\end{equation}
We thus obtain the integral kernel $K_A$, 
\begin{equation}
  K_A(x,y) = \tilde z^{-2} F(2,2-D/2, D/2+1; \tilde z^{-2}), \\
\end{equation}
and the integral kernel $K_B$~\cite{gusynin},
\begin{equation}
  K_B(x,y) = \dfrac{1}{\max(x,y)} F(1,2-D/2, D/2; \tilde z^{-2}).
\end{equation}
Using the Taylor expansion of the hypergeometric function
\begin{displaymath}
  F(\alpha,\beta,\gamma; z) = \sum_{\ell=0}^{\infty}
    \dfrac{(\alpha)_\ell (\beta)_\ell}{(\gamma)_{\ell}}
    \dfrac{z^\ell}{\ell!}, \qquad
  (\alpha)_\ell \equiv \alpha(\alpha+1)(\alpha+2)\cdots (\alpha+\ell-1),
\end{displaymath}
we obtain
\begin{eqnarray}
K_A(x,y)
       &\equiv & \frac{y}{x}\sum_{\ell=0}^{\delta/2} 
                 \frac{(-\delta/2)_\ell (\ell+1)}{(\delta/2+3)_\ell}
                 \left(\frac{y}{x}\right)^\ell \theta(x-y) +
                 (x \leftrightarrow y), \\
K_B(x,y)
       &\equiv & \frac{1}{x}\sum_{\ell=0}^{\delta/2}
                 \frac{(-\delta/2)_\ell}{(\delta/2+2)_\ell}
                 \left(\frac{y}{x}\right)^\ell \theta(x-y) +
                 (x \leftrightarrow y)
\label{K_B_imp}
\end{eqnarray}
for even dimensions $D=4+\delta \geq 4$.

\section{The NJL model in $D$ ($>4$) dimensions}
\label{app:NJL}

We consider the Nambu--Jona-Lasinio (NJL) model\footnote{
In order to avoid the complexity associated with the definition of 
continuous chiral symmetry in $D$ dimensions, we discuss in this Appendix
 the NJL model that has only discrete chiral symmetry 
 (it may be called the ``Gross-Neveu'' model.)}
in $D(>4)$ dimensions, 
\begin{equation}
  {\cal L}=\bar{\psi}i\Gamma^M \partial_M \psi +\frac{G}{2N}(\bar{\psi}\psi)^2.
\end{equation}
The gap equation is obtained from the self-consistency condition for the
dynamical mass $m$.
In the large $N$ limit, we find
\begin{eqnarray}
 m &=& \frac{\eta\, G \Lambda^{D-2}}{(4\pi)^{D/2}\Gamma(D/2)}
       \int_0^{\Lambda^2} dp_E^2 \,\left(\frac{p_E^2}{\Lambda^2}\right)^{D/2-1}
       \frac{m}{p_E^2+m^2}, 
   \nonumber\\
   &=& g \int_0^1 dz \,z^{D/2-1}
       \frac{m}{z+m^2/\Lambda^2},
\label{eq:NJLgap}
\end{eqnarray}
where $\eta\;(=2^{D/2})$ represents the dimension of the spinor
representation of $SO(1,D-1)$.
The dimensionless NJL coupling $g$ is defined by 
$g \equiv (4\pi)^{-D/2} \eta G \Lambda^{D-2}/\Gamma(D/2)$. 
Expanding the integrand of Eq.~(\ref{eq:NJLgap}) in terms of
$m/\Lambda$, we obtain
\begin{displaymath}
 \frac{1}{g} = \frac{1}{D/2-1}-\frac{1}{D/2-2}\frac{m^2}{\Lambda^2}+\cdots.
\end{displaymath}
The scaling behavior of the NJL model in $D\;(>4)$ dimensions is then
given by
\begin{equation}
 \frac{1}{g_{crit}}-\frac{1}{g} =
 \frac{2}{D-4}\frac{m^2}{\Lambda^2},
 \qquad
 g_{crit} \equiv D/2-1. 
\end{equation}
In order to obtain hierarchy between $m$ and $\Lambda$, we thus need
a fine tuning of the NJL coupling strength at the precision of the 
$(m/\Lambda)^2$ level {\it irrespectively of $D$ for $D>4$}.
The situation contrasts with the NJL model in dimensions less 
than four where the NJL coupling needs to be close to its critical
point at the precision of $(m/\Lambda)^{D-2}$.

\section{Approximate formulas for $g_{\rm eff}$}
\label{sec:app2}

\begin{figure}[t]
  \begin{center}
    \includegraphics{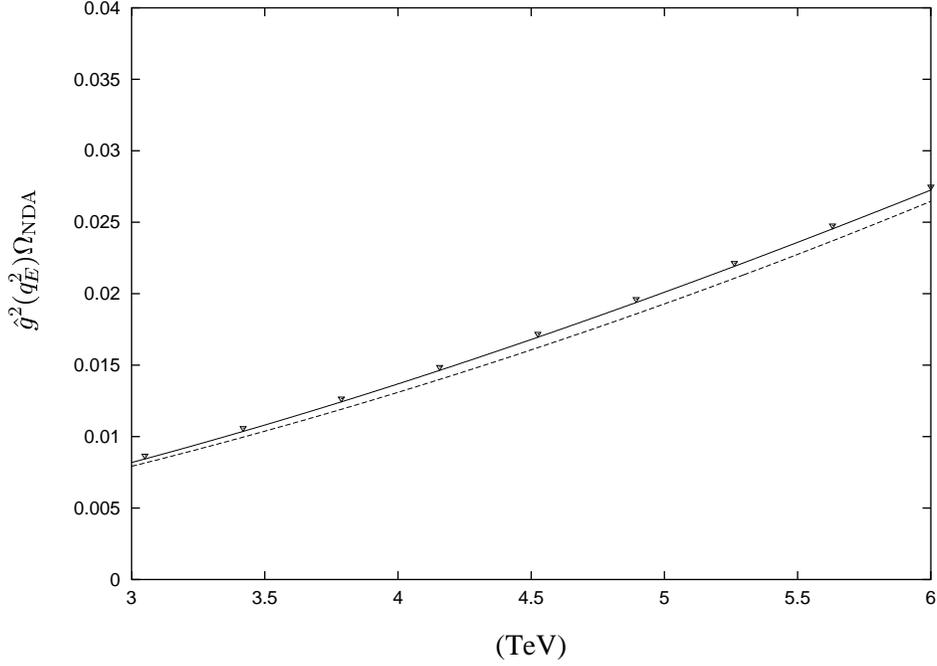}
    \caption{Graphs of the dimensionless gauge couplings.
             The solid line and the dashed line represent 
             the effective coupling of Eq.~(\ref{g_eff_K})
             and Eq.~(\ref{eq:sixdimeff}),
             respectively.  
             We also plot with the white triangles the coupling
             directly calculated from the definition of 
             Eq.~(\ref{eq:eff}) without using approximation
             Eq.~(\ref{eq:sum2int}).
             In this graph, we took $C_G=3$, $N_f=0$,
             $R^{-1}=3$ TeV, $\Lambda=6$ TeV.
             \label{fig:plot_alpha_sum}}
  \end{center}
\end{figure}

Equation (\ref{eq:eff}) can be further simplified by making several
approximations. 
We first concentrate our attention on $\Pi_{<}$.
This term depends on the renormalization scale $\mu$ and therefore it
can be minimized by taking an optimized choice of $\mu$. 
We assume that the appropriate $\mu^2$ is proportional to $-q^2$:
\begin{equation}
  \mu^2 = -\lambda q^2,
\end{equation}
with $\lambda$ being a constant which we will determine below.
Since the mass of the KK mode $m_{\vec n}$ is always lighter than the
renormalization scale $\mu$ in $\Pi_{<}$, we can safely neglect
$m_{\vec n}$ in the following analysis. 
It is straightforward to evaluate $\Pi_{<}$,
\begin{equation}
  (4\pi)^2 \Pi_{<}(q^2, 0 ; \mu = \sqrt{-\lambda q^2})
  = C_G \left[ 8  + \frac{4}{9}(2-D)\right]
    - \dfrac{5}{9} \eta T_R N_f
    - b' \ln \lambda. 
\label{eq:optla0}
\end{equation}
This term vanishes if we take the optimized value of $\lambda$:
\begin{equation}
  \ln\lambda 
   = \dfrac{ C_G\left(-8 -\frac{4}{9} (2-D)\right)
               +\frac{5}{9} \eta T_R N_f}
           {-b'}.
\label{eq:optla}
\end{equation}
We next turn to $\Pi_{>}$.  The KK mass is always heavier than the
renormalization scale $\mu$ in this term, while $\mu$ is proportional
to $q^2$ in the previous optimization procedure.
We therefore expand $\Pi_{>}$ in terms of the powers of 
$-q^2/m_{\vec n}^2$.
We find
\begin{equation}
  (4\pi)^2 \Pi_{>}(q^2, m^2) 
  = - C_G \left[
        \dfrac{2}{3} + \dfrac{2-D}{60}
      \right] \left(\dfrac{-q^2}{m^2}\right) 
    + \dfrac{\eta}{15} T_R N_f \left(\dfrac{-q^2}{m^2}\right) 
    + {\cal O}\left(\left(\dfrac{-q^2}{m^2}\right)^2 \right).
\label{eq:expand1}
\end{equation}
The sum of the KK modes can be approximated by replacing it with an
integral:
\begin{equation}
  \sum_{\vec k}^{m_{\vec k}>\mu} \rightarrow
    \frac{1}{n} \dfrac{2\pi^{\delta/2}}{\Gamma(\delta/2)} R
    \int_\mu^\Lambda dm (R m)^{\delta-1}, 
\label{eq:sum2int}
\end{equation}
which leads to 
\begin{equation}
  \sum_{\vec k}^{m_{\vec k}>\mu} \Pi_{>}(q^2, m_{\vec k}^2)
  \simeq \dfrac{\pi R^2}{(4\pi)^2 n} \left[
           -\frac{3}{5} C_G +\frac{\eta}{15} T_R N_f
         \right] (-q^2) \ln \dfrac{\Lambda^2}{\mu^2}
\label{eq:decouple}
\end{equation}
for  $D=4+\delta$, $\delta=2$.
Combining Eqs.~(\ref{eq:optla}), (\ref{eq:decouple}) and
(\ref{eq:effdimless}), it is now easy to obtain
Eq.~(\ref{eq:sixdimeff}).
The validity of Eq.~(\ref{eq:sixdimeff}) can be confirmed also 
numerically. (See 
Fig.\ref{fig:plot_alpha_sum}.) 
It should be noted that the approximation of
Eq.~(\ref{eq:expand1}) is not justified for $\lambda \ll 1$, however.
In order to obtain the upper bound of $\hat g_{\rm eff}$ including
such a possibility, we next evaluate the effective coupling without
use of the approximations of Eq.~(\ref{eq:optla0}) and
Eq.~(\ref{eq:expand1}). 
Using the approximation Eq.~(\ref{eq:sum2int}), we obtain a formula for 
the effective gauge coupling strength in six dimensions: 
\begin{eqnarray}
\dfrac{1}{\hat{g}_{\rm eff}^2 (q^2)} 
  &=& \dfrac{\mu^2}{(-q^2)} \left( 
         \dfrac{1}{\hat{g}_{\overline{MS}}^{2}(\mu)}+\frac{b'}{(4\pi)^3}
      \right) 
  \nonumber \\ 
  & & \quad 
    + \dfrac{1}{(4\pi)^3} \left[ 
       K_g(q^2,\Lambda^2)+K_b(q^2,\Lambda^2)+K_f(q^2,\Lambda^2)\right],
\label{g_eff_K}
\end{eqnarray}
with
\begin{eqnarray*}
K_g(q^2,\Lambda^2) 
  &\equiv& 4 C_G \left(
      \frac{5}{18}+\frac{1}{6} \ln \frac{\Lambda^2}{(-q^2)}
     +\frac{\Lambda^2}{(-q^2)} \tilde K_g(q^2, \Lambda^2)
  \right), \\
K_b(q^2,\Lambda^2) 
  &\equiv&  -2 C_G \left( 
      \frac{31}{450}+\frac{1}{30} \ln \frac{\Lambda^2}{(-q^2)}
     +\frac{\Lambda^2}{(-q^2)}\tilde K_b(q^2, \Lambda^2)
  \right), \\
K_f(q^2,\Lambda^2) 
  &\equiv& - 2\eta T_R N_f \left(
      \frac{47}{900}+\frac{1}{30} \ln \frac{\Lambda^2}{(-q^2)}
     +\frac{\Lambda^2}{(-q^2)}\tilde K_f(q^2, \Lambda^2)
  \right).
\end{eqnarray*}
The functions $\tilde K_g$, $\tilde K_b$, $\tilde K_f$ are defined by
\begin{eqnarray*}
  \tilde
  K_g(q^2, \Lambda^2) &\equiv& \int_0^1 dx f(q^2, \Lambda^2, x), \\
  \tilde
  K_b(q^2, \Lambda^2) &\equiv& \int_0^1 dx (2x-1)^2 f(q^2, \Lambda^2, x), \\
  \tilde
  K_f(q^2, \Lambda^2) &\equiv& \int_0^1 dx x(1-x) f(q^2, \Lambda^2, x),
\end{eqnarray*}
with 
\begin{displaymath}
 f(q^2,\Lambda^2,x) \equiv
   (1-x(1-x)\frac{q^2}{\Lambda^2})
   \ln\left(1-\dfrac{q^2}{\Lambda^2}x(1-x)\right).
\end{displaymath}
The integrals can be performed easily and we obtain (in the Euclidean
region $q^2<0$) 
\begin{eqnarray*}
\tilde K_g(q^2, \Lambda^2)
 &=& 
   -\frac{4}{3}+\frac{5}{18} \frac{q^2}{\Lambda^2}
   +\frac{1}{3}\left(4-\frac{q^2}{\Lambda^2}\right)^{3/2}
    \left(\dfrac{\Lambda^2}{-q^2}\right)^{1/2} 
    \tanh^{-1} \sqrt{\dfrac{-q^2}{4\Lambda^2-q^2}},
 \\
\tilde K_b(q^2, \Lambda^2)
 &=& 
   \frac{16}{15}\frac{\Lambda^2}{q^2} -\frac{28}{45}
  +\frac{31}{450}\frac{q^2}{\Lambda^2}
  +\frac{1}{15}\left(4-\dfrac{q^2}{\Lambda^2}\right)^{5/2}
   \left(\dfrac{\Lambda^2}{-q^2}\right)^{3/2}
   \tanh^{-1} \sqrt{\dfrac{-q^2}{4\Lambda^2-q^2}},
 \\
\tilde K_f(q^2, \Lambda^2)
  &=& 
  -\frac{4}{15}\dfrac{\Lambda^2}{q^2} -\frac{8}{45}
  +\frac{47}{900}\frac{q^2}{\Lambda^2}
  \\
  & &
  -\frac{1}{15}\left(4-\frac{q^2}{\Lambda^2}\right)^{3/2}
   \left(1+\dfrac{q^2}{\Lambda^2}\right)
   \left(\dfrac{\Lambda^2}{-q^2}\right)^{3/2}
   \tanh^{-1} \sqrt{\dfrac{-q^2}{4\Lambda^2-q^2}}.
\end{eqnarray*}
It is evident that $\hat g_{\rm eff}$ reaches its maximum at $q^2 =
-\Lambda^2$. (See Fig. \ref{fig:plot_alpha_sum}.)
We obtain
\begin{equation}
K \equiv \sum_{i=g,b,f}K_i(-\Lambda^2,\Lambda^2)
  = C_G\left(
     -\frac{88}{45}
     +\dfrac{10\sqrt{5}}{3}\tanh^{-1}\dfrac{1}{\sqrt{5}}
    \right)
   -\dfrac{8\eta T_R}{45} N_f.
\label{eq:eff_UV}
\end{equation}
Equation (\ref{g_eff_K}) thus leads to an upper bound on $\hat g_{\rm eff}$,
\begin{equation}
  \hat g_{\rm eff}^2(q^2) < \dfrac{(4\pi)^3}{K} \qquad
  \mbox{for $0\le -q^2 \le \Lambda^2$},
\label{eq:upper_bound}
\end{equation}
where we have assumed that the $\overline{MS}$ coupling is below its 
UV fixed point, 
\begin{equation}
   \hat g_{\overline{MS}}^2 < g_*^2 =\dfrac{(4\pi)^3}{-b'}.
\end{equation}
It should be emphasized that Eq.~(\ref{eq:eff_UV}) is independent of
the cutoff $\Lambda$
and thus
the upper bound Eq.~(\ref{eq:upper_bound}) can be adopted for arbitrary 
$\Lambda$.
In order to clarify the point,
it is illuminating to define $\hat g_\Lambda$ by
\begin{equation}
  \hat g_\Lambda^2 \equiv \hat g_{\rm eff}^2(q^2 = - \Lambda^2).
\end{equation}
The coupling $\hat g_\Lambda$ can be regarded as a ``bare parameter''
of the present model.
An analogue of the ``$\beta$ function'' for $\hat g_\Lambda$ is then
given by
\begin{equation}
\beta(\hat g_\Lambda) \equiv
 \Lambda\frac{d}{d\Lambda} \hat g_\Lambda
  = \hat g_\Lambda - \dfrac{K}{(4\pi)^3} \hat g_\Lambda^3.
  \label{eq:betalambda}
\end{equation}
The upper limit $(4\pi)^3/K$ is thus given by the UV fixed point of
the ``$\beta$ function'' Eq.~(\ref{eq:betalambda}).

Numerically we obtain
$K \simeq 1.63 C_G - 0.71 T_R N_f$.
For large $-b'$ or large $C_G$, we thus find that the upper bound is
roughly proportional to the UV fixed point in the $\overline{MS}$ scheme,
\begin{equation}
  \dfrac{(4\pi)^3}{K} \simeq 2 \dfrac{(4\pi)^3}{-b'} = 2 g_*^2.
\end{equation}

\end{document}